\documentclass[twocolumn,english,reprint,superscriptaddress,longbibliography,pra]{revtex4-1}

\usepackage[T1]{fontenc}
\usepackage[latin9]{inputenc}
\usepackage{color}
\usepackage{bm}
\usepackage{amstext}
\usepackage{amssymb}
\usepackage{graphicx}
\usepackage{booktabs}
\usepackage{tabularx}
\usepackage{amsmath,mathrsfs}
\usepackage{enumitem}
\makeatletter
\usepackage{multibib}
\DeclareMathOperator*{\argmin}{argmin}   

\newcommand{\SM}{[\hyperref[SM]{SM}]}
\usepackage{babel}
\usepackage{bbold}
\usepackage{hyperref}
\hypersetup{
 linktocpage = true,
     colorlinks,
    citecolor=blue,
    filecolor=black,
    linkcolor=blue,
    urlcolor=blue
}

\usepackage{wasysym}

\makeatother

\begin{document}
\author{Giacomo Torlai}
\email{gtorlai@flatironinstitute.org}
\affiliation{Center for Computational Quantum Physics, Flatiron Institute, New York, NY 10010, USA}

\author{Guglielmo Mazzola}
\affiliation{IBM Research Zurich, Saumerstrasse 4, 8803 Ruschlikon, Switzerland}

\author{Giuseppe Carleo}
\affiliation{Center for Computational Quantum Physics, Flatiron Institute, New York, NY 10010, USA}

\author{Antonio Mezzacapo}
\affiliation{IBM T.J. Watson Research Center, Yorktown Heights, NY 10598, USA}

\title{Precise measurement of quantum observables with neural-network estimators}

\begin{abstract}
The measurement precision of modern quantum simulators is intrinsically constrained by the limited set of measurements that can be efficiently implemented on hardware. 
This fundamental limitation is particularly severe for quantum algorithms where complex quantum observables are to be precisely evaluated. To achieve precise estimates with current methods, prohibitively large amounts of sample statistics are required in experiments. Here, we propose to reduce the measurement overhead by integrating artificial neural networks with quantum simulation platforms. We show that unsupervised learning of single-qubit data allows the trained networks to accommodate measurements of complex observables, otherwise costly using traditional post-processing techniques. The effectiveness of this hybrid measurement protocol is demonstrated for quantum chemistry Hamiltonians using both synthetic and experimental data. Neural-network estimators attain high-precision measurements with a drastic reduction in the amount of sample statistics, without requiring additional quantum resources.

\end{abstract}

\maketitle


The measurement process in quantum mechanics has far-reaching implications, ranging from the fundamental interpretation of quantum theory~\cite{schlosshauer2013snapshot} to the design of quantum hardware~\cite{clerk2010introduction}. The advent of medium-sized quantum computers has drawn attention to scalability issues different than control errors or decoherence, which nonetheless hinder the realization of complex quantum algorithms. Coherent and incoherent noise altering quantum states can be corrected in fault-tolerant architectures~\cite{Campbell2017}. In contrast, the fluctuations introduced by a non-ideal measurement protocol lead to intrinsic quantum noise which persists even in a fault-tolerant regime.

The most promising quantum computing platforms, such as superconducting or ion-trap processors, provide access to projective single-qubit non-demolition measurements~\cite{wallraff2004strong, wineland1980double}. Armed with these simple measurements, one is faced with a plethora of quantum simulation algorithms which rely on accurate estimations of specialized observables. For practical purposes, in order to suppress the uncertainty arising from a sub-optimal measurement apparatus, massive amounts of sample statistics need to be generated by the quantum device~\cite{wecker2015progress}. Complex estimators are then reconstructed through classical post-processing of single-qubit data.

As the measurement precision remains tied to the interface between the quantum and the classical hardware, it becomes critical to develop methods capable of extracting more information from a given measurement dataset~\cite{Jena2019,Yen2019b,Huggins2019,Gokhale2019,Crawford2019,zhao2019measurement}.
Given this, data-driven algorithms can provide a viable path towards improved accuracy and scalability in quantum simulation platforms.

\begin{figure}[b!]
\noindent \centering \includegraphics[width=0.85\columnwidth]{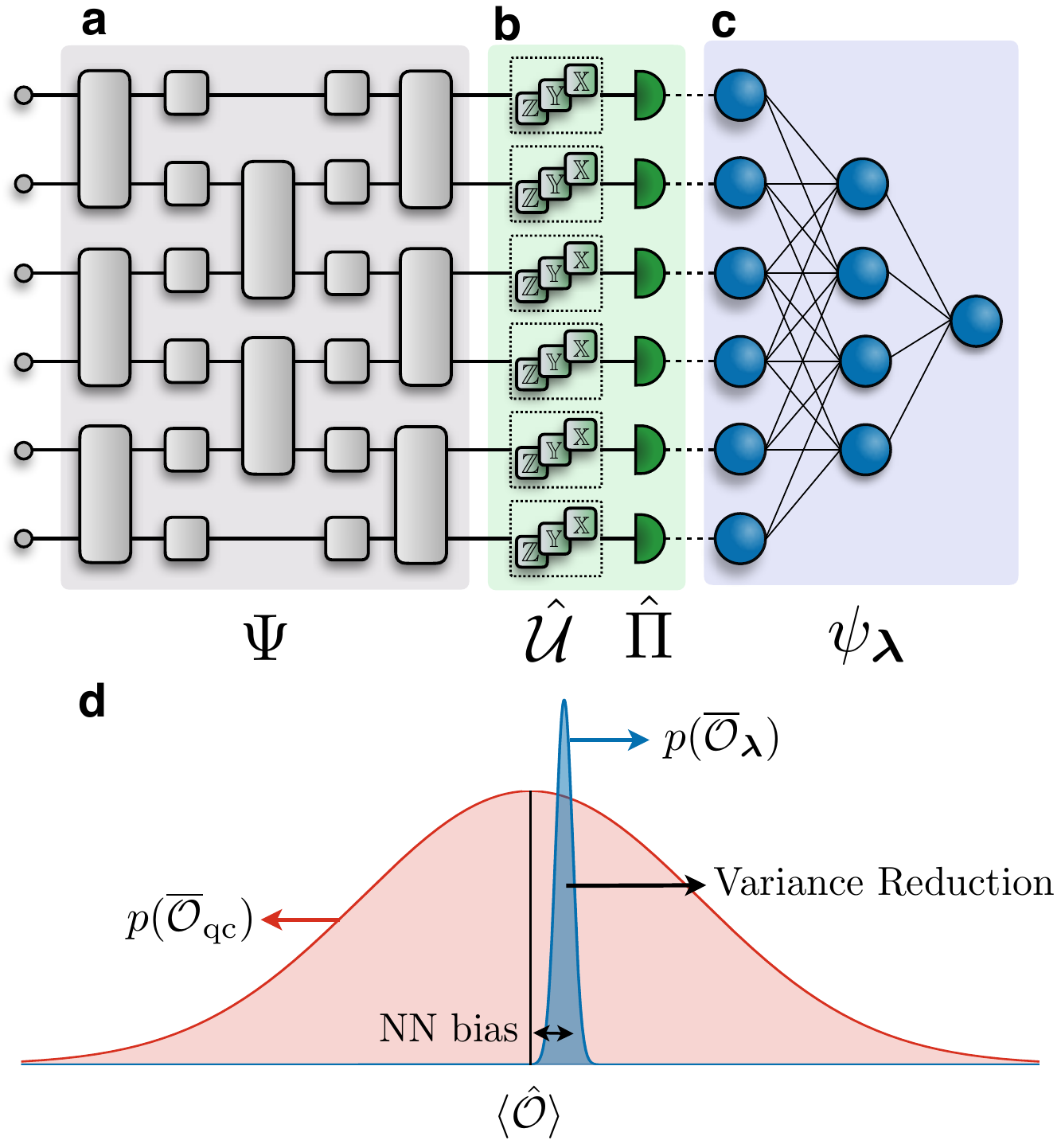}
\caption{Measurements on quantum hardware with neural-network estimators. ({\bf a}) A quantum circuit prepares a quantum state $\Psi$. ({\bf b}) Single-qubit measurements, consisting of a local rotation $\hat{\mathcal{U}}$ and a projective measurement $\hat{\Pi}$. ({\bf c}) A neural network is trained on the output of the measuring apparatus to discover a representation $\psi_{\bm{\lambda}}$ of the state $\Psi$ that retrieves the expectation value of a target quantum observable $\hat{\mathcal{O}}$. ({\bf d}) The intrinsic measurement uncertainty is traded for a systematic reconstruction bias, leading to a measurement outcome distribution with lower variance.}
\label{Fig::1}
\end{figure}

Machine learning has recently shown its flexibility in finding approximate solutions to complex problems in a broad range of physics~\cite{MLQreview}. In particular, extensive theoretical work has demonstrated the potential of artificial neural networks in the context of quantum many-body physics~\cite{Carrasquilla17,LeiWang16,Carleo17,Torlai16,Evert17,neural_rg,torlai_2018_nnqst,bukov18}. The same approach has also been employed to enhance the capabilities of various quantum simulation platforms~\cite{Seif_2018,Rem_MLcoldatoms,Bohrdt_MLfermions,torlai_rydberg,Zhang_MLcuprates,islam2019}. With the increasing stream of quantum data produced in laboratories, it is natural to expect further synergy between machine learning and experimental quantum hardware.

In this Article, we propose to integrate neural networks with quantum simulators to increase the measurement precision of quantum observables. Using unsupervised learning on single-qubit data to learn approximately the quantum state underlying the hardware, neural networks can be deployed to generate estimators free of intrinsic quantum noise. This comes at a cost of a systematic bias from the imperfect quantum state reconstruction. We investigate the trade-off between these two sources of uncertainty for measurements of quantum chemistry Hamiltonians, costly with standard techniques~\cite{wecker2015progress}. We show a reduction of various orders of magnitude in the amount of data required to reach chemical accuracy for simulated data. For experimental data produced by a superconducting quantum hardware, we recover energy estimates with a low number of data points. 
This opens new opportunities for quantum simulation on near-term quantum hardware~\cite{Preskill2018}.


\section*{Neural-network estimators}
We examine the task of estimating the expectation value of a generic observable $\hat{\mathcal{O}}$ on a quantum state $|\Psi\rangle$ prepared by a quantum computer with $N$ qubits. A direct measurements produces an estimator $\overline{\mathcal{O}}\approx\langle\hat{\mathcal{O}}\rangle$ with sample variance $\sigma^2[\mathcal{O}]\approx\langle{\hat{\mathcal{O}}}^2\rangle-\langle{\hat{\mathcal{O}}}\rangle^2$. This measurement is optimal when $|\Psi\rangle$ is eigen-state of $\hat{\mathcal{O}}$ (i.e. $\sigma^2[\mathcal{O}]=0$), but requires sample statistics from the observable eigen-basis, typically not available on a quantum computer. 

A more flexible measurement protocol can be devised by considering the expansion of the observable $\hat{\mathcal{O}}$ in terms of $K$ tensor products of Pauli operators
\begin{equation}
\hat{\mathcal{O}}=\sum_{k=1}^Kc_k\hat{P}_k\quad,\quad \hat{P}_k\in\{\hat{\mathbb{1}},\hat{\sigma}^x,\hat{\sigma}^y,\hat{\sigma}^z\}^{\otimes N}\:,
\label{Eq::Obs}
\end{equation} 
where $c_k$ are real coefficients. This decomposition allows one to estimate the expectation value from independent measurements of each Pauli operator, only requiring single-qubit data. In contrast to the direct measurement, the final estimator $\overline{\mathcal{O}}_{\text{qc}}$ suffers an increased uncertainty $\epsilon_{\text{qc}}=\sqrt{\sum_k |c_k|^2\sigma^2[P_k]/S}$, where $\sigma^2[P_k]$ is the sample variance of $\hat{P}_k$ and $S$ is the number of measurements (see Supplementary Material \SM). The overhead in sample statistics to reduce this uncertainty becomes particularly severe for observables with a large number $K$ of Pauli operators.

We overcome this limitation by deploying unsupervised machine learning on single-qubit data to obtain an approximate reconstruction of the quantum state $|\Psi\rangle$ (Fig.~\ref{Fig::1}). We call this reconstruction approximate in the sense that, unlike traditional quantum state tomography~\cite{Banaszek2013}, we are primarily interested in the more restricted task of recovering measurement outcomes for the observable $\hat{\mathcal{O}}$. We first parametrize a generic many-body wavefunction by an artificial neural network. In a given reference basis $|\bm{\sigma}\rangle$ of the many-body Hilbert space (e.g. $|\bm{\sigma}\rangle=|\sigma_1^z,\dots,\sigma_N^z\rangle$, $\sigma_i^{z} = \{0,1\}$), the neural network provides a parametric encoding of the amplitudes $\psi_{\bm{\lambda}}(\bm{\sigma})=\langle\bm{\sigma}|\psi_{\bm{\lambda}}\rangle$ into a set of complex-valued weights $\bm{\lambda}$~\cite{Carleo17}. Specifically, we implement the restricted Boltzmann machine (RBM)~\cite{Hinton85}, a physics-inspired generative neural network currently explored in many areas of condensed matter physics and quantum information~\cite{mlNISQ,RBM_natphys}. 

\begin{figure}[t]
\noindent \centering \includegraphics[width=0.9\columnwidth]{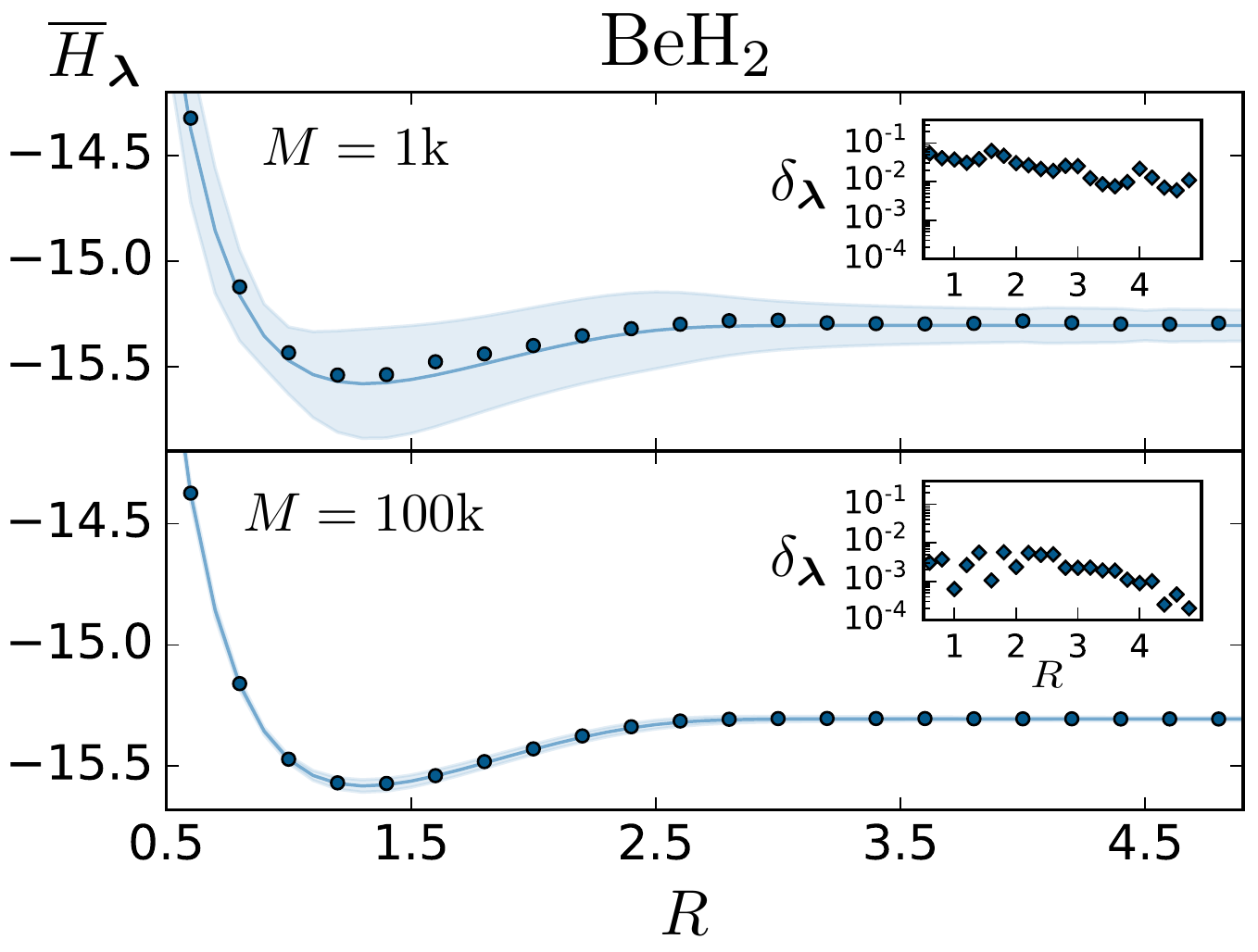}
\caption{Reconstruction of the potential energy surface of the BeH$_2$ molecule (Hartree and Angstrom units). We show, for different dataset sizes $M$, the comparison between the exact ground state energy $E_0$ (solid lines) and the energies obtained with the neural-network estimator (markers). The shaded regions span one standard deviation for the estimate on the quantum hardware with standard averaging method using $M$ measurements. In the insets, we show the deviations $\delta_{\bm{\lambda}}=|E_0-\overline{H}_{\bm{\lambda}}|$ of the RBM estimators from the exact energies.}
\label{Fig::2}
\end{figure}

\begin{figure*}[t!]
\noindent \centering \includegraphics[width=2\columnwidth]{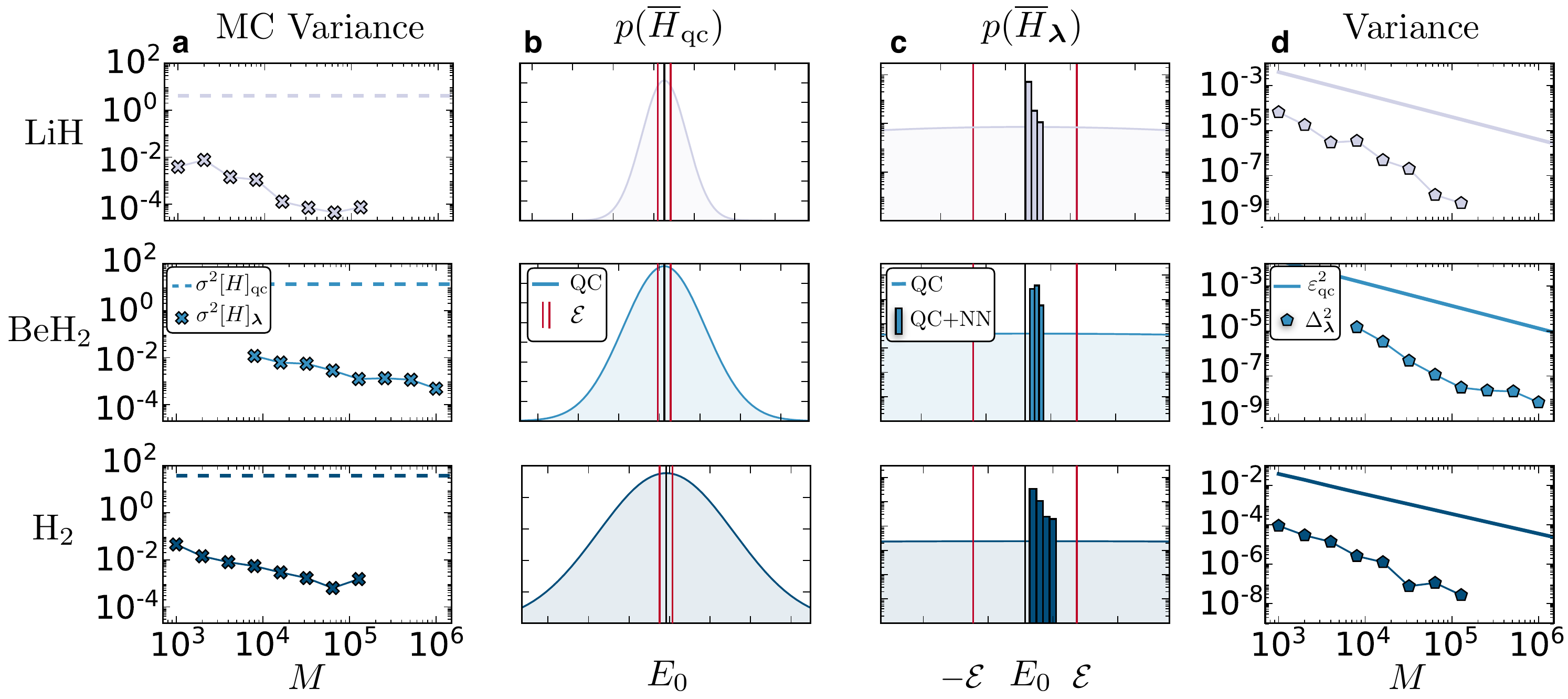}
\caption{Measurement uncertainty of neural-network estimators for molecular systems. The energy units are Hartrees. ({\bf a}) Statistical variance in the MC sampling of the neural network as a function of the size $M$ of the training dataset. We compare this with the variance $\sigma^2[H]_{\text{qc}}= \sum_{k=1}^K |c_k|^2 \sigma^2[P_k]$ arising from independent estimation of each Pauli operator with $S=M/K$ measurements. ({\bf b}) Energy measurement distribution on the quantum computer. The red lines bounds the chemical accuracy interval $\mathcal{E}=1.6\times10^{-3}$. The total number of measurements $M$ is set to 64k, 512k, 64k for the LiH, BeH$_2$ (STO-3G basis) and H$_2$ (6-31G basis) molecules respectively. ({\bf c}) Energy measurement distribution of the neural-network estimator (histogram), on the same number of measurements used in ({\bf b}). All estimates fall within chemical accuracy from the true value $E_0$. ({\bf d}) Energy errors induced by imperfect state reconstruction. We compare, for different dataset sizes $M$,  the sample variance of the mean $\varepsilon_{\text{qc}}^2=\sigma^2[H]_{\text{qc}}/S$ with the variance of the distribution of the neural-network estimator $\Delta^2_{\bm{\lambda}}$, calculated from the energy histograms in ({\bf c}).}
\label{Fig::3}
\end{figure*}

The quantum state reconstruction is carried out by training the neural network on a dataset $\mathcal{D}$ of $M$ single-qubit projective measurements, obtained from the target quantum state $|\Psi\rangle$ prepared by the hardware. Using an extension of unsupervised learning~\cite{torlai_2018_nnqst}, the network parameters $\bm{\lambda}$ are optimized via gradient descent to minimize the statistical distance between the probability distribution underlying the data and the RBM projective measurement probability. We adopt the standard measure given by the Kullbach-Leibler divergence
\begin{equation}
\mathcal{C}_{\bm{\lambda}}=-\frac{1}{M}\sum_{\bm\sigma^{\bm{b}} \in \mathcal{D}}\log|\psi_{\bm{\lambda}}( \bm\sigma^{\bm{b}})|^2\:,
\end{equation} 
where $\bm\sigma^{\bm{b}}$ is a $N$-bit string $(\sigma_1^{b_1},\dots,\sigma_N^{b_N})$ and $\bm{b}$ are Pauli bases ($b_j=\{x,y,z\}$) drawn uniformly from the set of Pauli operators $P_k$ appearing in Eq. (\ref{Eq::Obs}) \SM.

Once the optimal parameters are selected according to cross-validation on held-out data, measurements of specialized observables can be performed by the neural network~\cite{Torlai16}. The expectation value of the quantum observable is simply approximated by the statistical estimator
\begin{equation}
\overline{\mathcal{O}}_{\bm{\lambda}}=\frac{1}{n_{\text{mc}}}\sum_{j=1}^{n_{\text{mc}}}\frac{\langle\bm{\sigma}_j|\hat{\mathcal{O}}|\psi_{\bm{\lambda}}\rangle}{\langle\bm{\sigma}_j|\psi_{\bm{\lambda}}\rangle}\:,
\label{Eq::RBM_obs}
\end{equation}
where $\{\bm{\sigma}_1,\dots,\bm{\sigma}_{n_{\text{mc}}}\}$ is a set of $n_{\text{mc}}$ configurations drawn from the probability distribution $|\psi_{\bm{\lambda}}(\bm{\sigma})|^2$ via Monte Carlo (MC) sampling. Here lies the critical advantage of the neural-network estimator: despite that the wavefunction $\psi_{\bm{\lambda}}(\bm{\sigma})$ is reconstructed from single-qubit data generated by the quantum computer, the measurement it produces is not affected by the intrinsic quantum noise. This is in fact equivalent to the direct measurement scheme where data is collected in the eigen-basis of the observable $\hat{\mathcal{O}}$ \SM.

\section*{Results}
We benchmark our technique on molecular Hamiltonians $\hat{H}$, an exemplary test cases for complex observables. For these fermionic systems, the number of Pauli operators $K$ in Eq.~(\ref{Eq::Obs}) can grow up to the fourth power in the number of orbitals considered~\cite{DBLP:journals/qic/PoulinHWWDT15}. The resulting fast growth in measurement complexity remains a roadblock for quantum simulations on near-term hardware based on low depth quantum-classical hybrid algorithms, such as variational quantum eigensolvers~\cite{Kandala17}.

We generate synthetic measurement datasets, sampling from the exact ground state of Beryllium Hydrate (BeH$_2$), calculated by exact diagonalization of a $N=6$ qubit Hamiltonian. The latter is obtained from a second-quantized fermionic Hamiltonian in the atomic STO-3G basis through a parity transformation and qubit tapering from molecular symmetries~\cite{Bravyi17,Kandala17}.
We train a set of RBMs at different inter-atomic separations $R$ using datasets $\mathcal{D}$ of increasing size $M$, and perform measurements of the molecular Hamiltonians $\overline{H}_{\bm{\lambda}}$. We show in Fig.~\ref{Fig::2} the neural-network estimators over the entire molecular energy surface. Comparing these measurements with exact energies shows that a relatively good precision can be achieved with as low as $M=10^3$ (total) measurements. For a given number of measurements $M$, the neural-network estimator provides better estimates with respect to the conventional estimator $\overline{H}_{\text{qc}}$.

The higher precision of estimators produced by the neural networks originates from the direct parametrization of the many-body wavefunction, eliminating any intrinsic quantum noise. In turn, the imperfect quantum reconstruction leads to two additional sources of uncertainty: a MC variance of statistical nature, and a systematic bias in the expectation value. In the following, we investigate these noise sources for the BeH$_2$ molecule, as well as the Lithium Hydrate (LiH) in STO-3G basis and the Hydrogen (H$_2$) molecule in the 6-31G basis, encoded in $N=4$ and $N=8$ qubits respectively. We estimate the uncertainty of the measurement with the quantum computer using the exact variance calculated on the ground state wavefunction, with $S=M/K$ measurements per Pauli operator. For all molecules, we consider the geometry at the bond distance.

The statistical uncertainty from the MC averaging is given by $\epsilon_{\bm{\lambda}}=\sqrt{\sigma^2[H]_{\bm{\lambda}}/n_{\text{mc}}}$, where $\sigma^2[H]_{\bm{\lambda}}$ is the variance of $\hat{H}$ on the samples generated by the neural network \SM. Since the target state (i.e. ground state) is an eigenstate of the observable, a perfect reconstruction would lead to zero variance. Deviations from the exact ground state set the amount $\sigma^2[H_{\bm{\lambda}}]>0$ of statistical uncertainty in the sampling. In Fig.~\ref{Fig::3}a, we show the MC variance for training datasets of increasing size $M$ . Here, we fix the amount of MC samples to $n_{\text{mc}}=10^5$, which is sufficient to make statistical fluctuations negligible. As expected, the MC variance decreases as $M$ grows larger and the quality of the reconstruction improves, with significant reduction compared to the variance $\sigma^2[H]_{\text{qc}}$ obtained from standard post-processing.

\begin{figure}[t]
\noindent \centering \includegraphics[width=0.9\columnwidth]{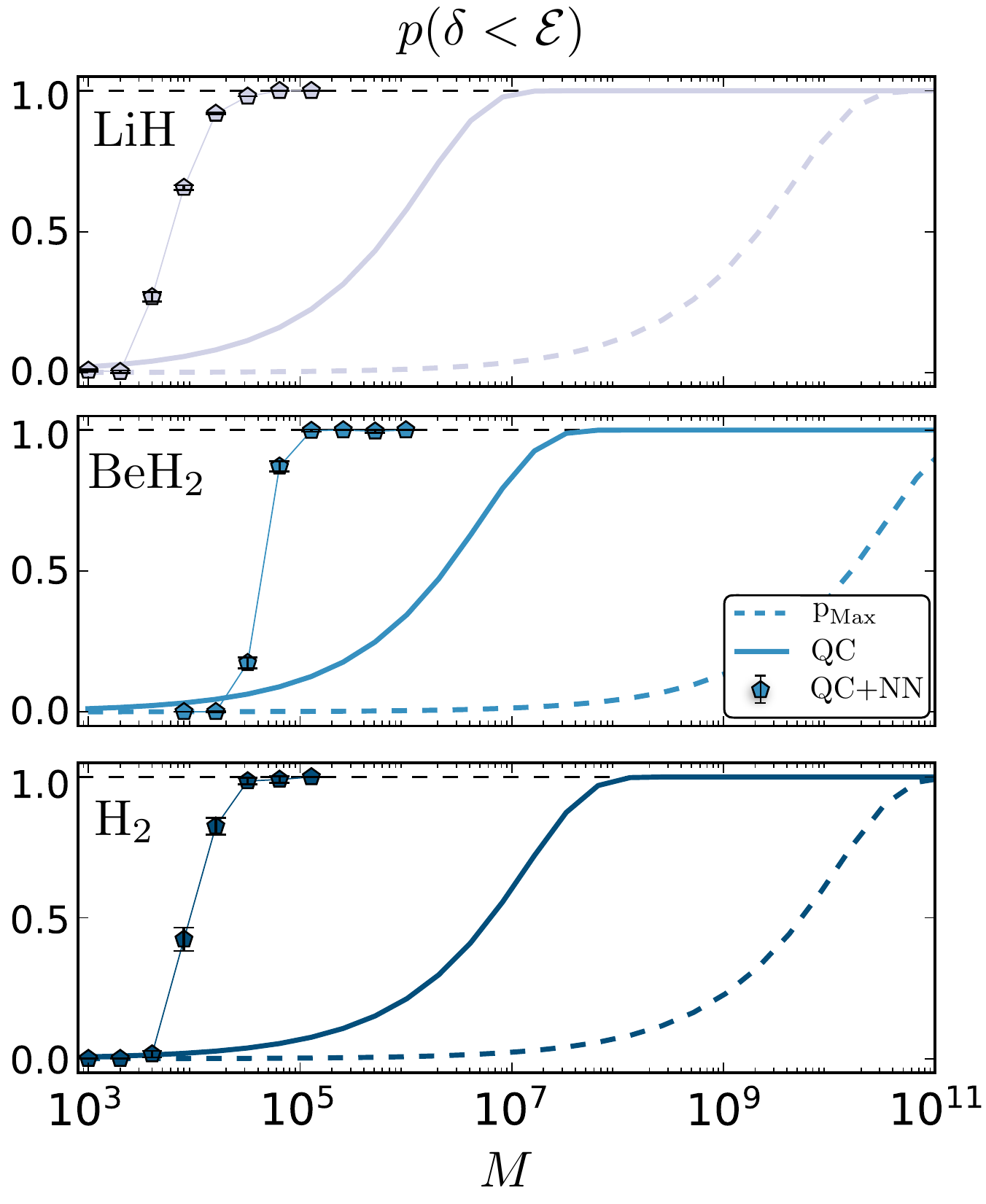}
\caption{Probability of reaching chemical accuracy as a function of the total number of measurements $M$. We compute the probability $p(\delta<\mathcal{E})$ to obtain a final energy estimate within chemical accuracy from the exact ground state energy between. Plotted are probabilities for the neural-network estimator and standard one. An upper bound $p_{\text{Max}}$ to the latter is also shown, obtained by setting $\sigma^2[H]_{\text{qc}}=(\sum_k|c_k|)^2$.}
\label{Fig::4}
\end{figure}

The reconstruction error in the neural-network estimator is also affected by finite-size deviations in the training dataset. To understand this contribution, we train a collection of 100 RBMs on independent measurement datasets, and compare the measurement distribution with the one obtained from standard averaging (Fig.~\ref{Fig::3}b). By examining histograms of energies built from separate dataset realizations  (Fig.~\ref{Fig::3}c), we observe that for sufficiently large $M$ the distribution of the neural-network estimator sharply peaks and gets close to the exact expectation value, with a positive off-set due to the energy variational principle. For a quantitative comparison between the two distributions, we show in Fig.~\ref{Fig::3}d the variance of the mean for the neural-network estimator $\epsilon^2_{\bm \lambda}$ (estimated from the histograms). We observe about two orders of magnitude improvement over the uncertainty $\epsilon^2_{\text{qc}}$ of the standard estimator. Further systematic errors due to approximate representability have been shown to be negligible for molecular systems of larger sizes~\cite{choo_fermionic_2019}.

The total uncertainty in the final measurement estimator is a combination of systematic bias and statistical noise. We quantify the combined effect by considering the probability $p(\delta<\mathcal{E})$ that the deviation $\delta=|E_0-\overline{H}|$ from the ground state energy $E_0$ is smaller than chemical accuracy $\mathcal{E}$. The specific value, which depends on thermal fluctuations at room temperature, is fixed to $\mathcal{E}=1.6\times10^{-3}$ Ha. A simple calculation leads to $p(\delta_{\text{qc}}<\mathcal{E}) = \text{Erf}(\:\mathcal{E}\:\sqrt{S/2\sigma^2[H]_{\text{qc}}})$ for the standard estimator. We evaluate this probability for the neural-network estimator by independently re-sampling each neural-network across the separate training realizations. We show the results in Fig.~\ref{Fig::4}, where we also include an upper bound $p_{\text{Max}}$ often referenced in literature~\cite{wecker2015progress, Huggins2019}. We observe drastic improvements up to three orders of magnitude in the total measurements $M$ required to get to $p(\delta_{\bm{\lambda}}<\mathcal{E})=1$.

\begin{figure}[t]
\noindent \centering \includegraphics[width=0.9\columnwidth]{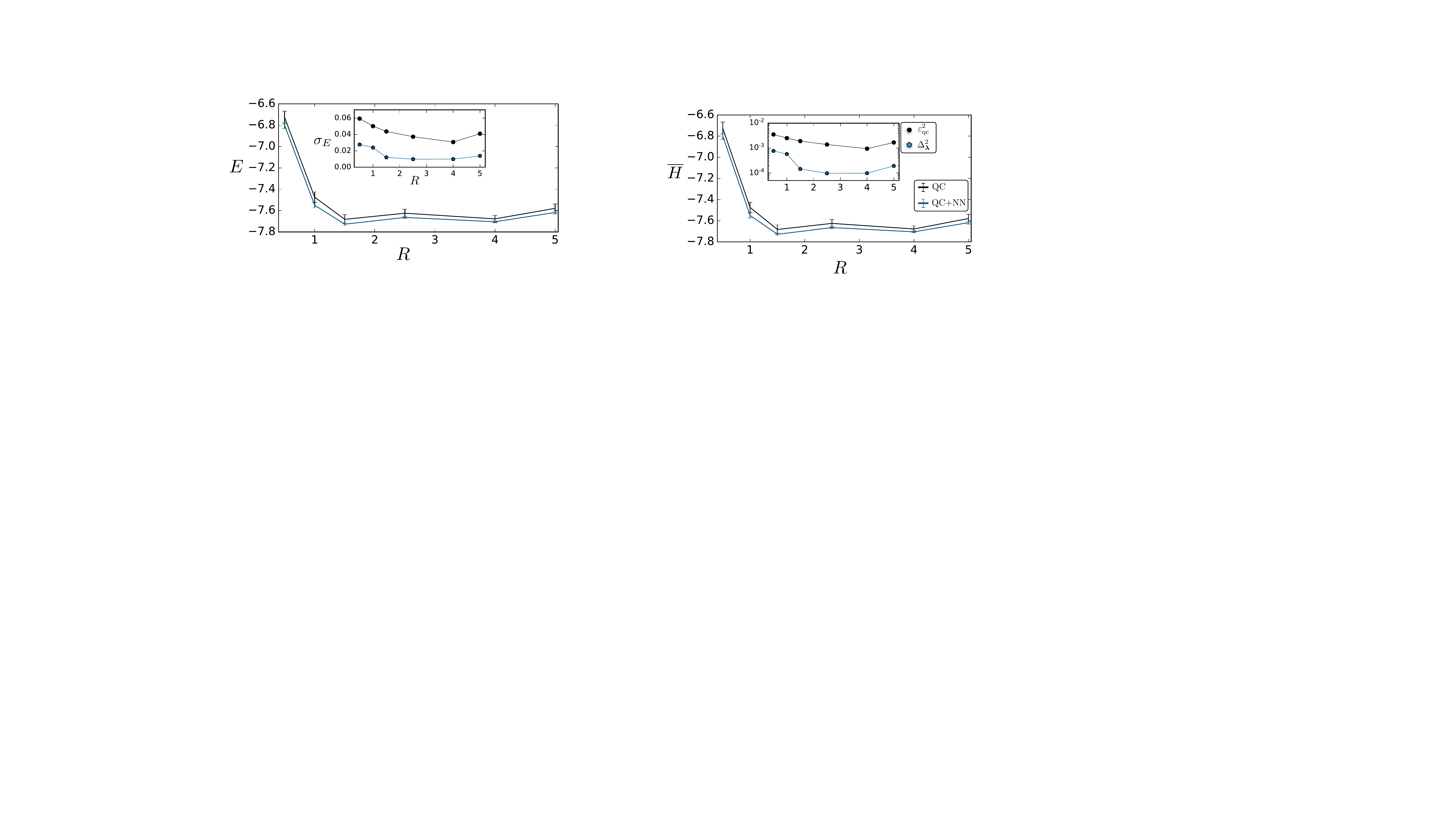}
\caption{Molecular energy (Hartrees) of LiH from experimental data generated by a superconducting quantum processor~\cite{Kandala19}, as function of the interatomic distance (Angstroms). The inset shows the variance $\varepsilon^2_{\text{qc}}$ from sub-sampling $5\times 10^3$ measurements out of $2.5\times 10^6$ data points, and the corresponding variance $\Delta^2_{\bm{\lambda}}$ obtained by separate neural-network reconstructions.}
\label{Fig::5}
\end{figure}

Finally, we show estimations of molecular energies with experimental data obtained in a variational quantum eigensolver. We use data from Ref.~\cite{Kandala19}, which consist of samples from an approximate ground state preparation of the LiH molecule on superconducting quantum hardware. In Fig.~\ref{Fig::5}, we plot the energy profile reconstructed by the neural network, showing a good agreement using only a fraction of the total experimental measurements. Note that decoherence determines a discrepancy between the reconstructed and the measured profile, since our RBM makes a pure state assumption, which is not exactly verified experimentally. 

To estimate the uncertainty, we train 50 RBMs on separate datasets obtained by sub-sampling $M=5\times10^3$ measurement data points, out of the original $2.5\times10^6$ measurements in ~\cite{Kandala19}. Despite the mixing in the quantum state underlying the measurements, the uncertainties in the neural-network estimators are systematically lower than the standard measurement scheme, similarly to what observed in synthetic data.

\section*{Discussion}
We have introduced a novel procedure to measure complex observables in quantum hardware. The approach is based on approximate quantum state reconstruction tailored to retrieve a quantum observable of interest. 
For the particularly demanding case of quantum chemistry applications, we have provided evidence that neural-network estimators achieve precise measurements with a reduced amount of sample statistics. 

An intriguing open question for future research is the systematic understanding of machine learning-based quantum state reconstruction. 
For positive wavefunctions, a favorable asymptotic reconstruction scaling has been recently shown~\cite{Sehayek2019}. For non-positive states, such as ground states of interacting electrons, recent works addressed representability~\cite{backflowNN,choo_fermionic_2019,deepmindfermions}, while much less is known for the reconstruction complexity, leaving open prospects for future studies. 

For measurement data generated by the experimental hardware, we have also assumed that the quantum state is approximately pure. When decoherence effects substantially corrupt the state, density-matrix neural-network reconstruction techniques~\cite{torlai_ndo,carrasquilla_povm} could be employed as an alternative to the algorithm presented here. Generative modelsother than neural networks~\cite{glasser2019expressive}  could also be explored in this setup.

Finally, the increased measurement precision with lower sample complexity makes neural-network estimators a powerful asset in variational quantum simulations of ground states by hybrid algorithms using low-depth quantum circuits~\cite{Kandala17,zoller19}. It is natural to expect integration of generative models in the feedback loop for the quantum circuit optimization. With the ever increasing size of quantum hardware, we envision that machine learning will play a fundamental role in the development of the next generation of quantum technologies.

\section*{Acknowledgements}
We thank J. Carrasquilla, M. T. Fishman, J. Gambetta and R. G. Melko for useful discussions. We thank A. Kandala for the availability of raw experimental data from Ref.~\cite{Kandala19}. The Flatiron Institute is supported by the Simons Foundation. A.M. acknowledges support from the IBM Research Frontiers Institute. Numerical simulations have been carried out on the Simons Foundation Supercomputing Center. Quantum state reconstruction was performed with the NetKet software~\cite{netket}, and the molecular Hamiltonians were obtained with Qiskit Aqua~\cite{qiskit}.

\bibliography{bibliography.bib}

\begin{thebibliography}{46}%
\makeatletter
\providecommand \@ifxundefined [1]{%
 \@ifx{#1\undefined}
}%
\providecommand \@ifnum [1]{%
 \ifnum #1\expandafter \@firstoftwo
 \else \expandafter \@secondoftwo
 \fi
}%
\providecommand \@ifx [1]{%
 \ifx #1\expandafter \@firstoftwo
 \else \expandafter \@secondoftwo
 \fi
}%
\providecommand \natexlab [1]{#1}%
\providecommand \enquote  [1]{``#1''}%
\providecommand \bibnamefont  [1]{#1}%
\providecommand \bibfnamefont [1]{#1}%
\providecommand \citenamefont [1]{#1}%
\providecommand \href@noop [0]{\@secondoftwo}%
\providecommand \href [0]{\begingroup \@sanitize@url \@href}%
\providecommand \@href[1]{\@@startlink{#1}\@@href}%
\providecommand \@@href[1]{\endgroup#1\@@endlink}%
\providecommand \@sanitize@url [0]{\catcode `\\12\catcode `\$12\catcode
  `\&12\catcode `\#12\catcode `\^12\catcode `\_12\catcode `\%12\relax}%
\providecommand \@@startlink[1]{}%
\providecommand \@@endlink[0]{}%
\providecommand \url  [0]{\begingroup\@sanitize@url \@url }%
\providecommand \@url [1]{\endgroup\@href {#1}{\urlprefix }}%
\providecommand \urlprefix  [0]{URL }%
\providecommand \Eprint [0]{\href }%
\providecommand \doibase [0]{http://dx.doi.org/}%
\providecommand \selectlanguage [0]{\@gobble}%
\providecommand \bibinfo  [0]{\@secondoftwo}%
\providecommand \bibfield  [0]{\@secondoftwo}%
\providecommand \translation [1]{[#1]}%
\providecommand \BibitemOpen [0]{}%
\providecommand \bibitemStop [0]{}%
\providecommand \bibitemNoStop [0]{.\EOS\space}%
\providecommand \EOS [0]{\spacefactor3000\relax}%
\providecommand \BibitemShut  [1]{\csname bibitem#1\endcsname}%
\let\auto@bib@innerbib\@empty
\bibitem [{\citenamefont {Schlosshauer}\ \emph {et~al.}(2013)\citenamefont
  {Schlosshauer}, \citenamefont {Kofler},\ and\ \citenamefont
  {Zeilinger}}]{schlosshauer2013snapshot}%
  \BibitemOpen
  \bibfield  {author} {\bibinfo {author} {\bibfnamefont {Maximilian}\
  \bibnamefont {Schlosshauer}}, \bibinfo {author} {\bibfnamefont {Johannes}\
  \bibnamefont {Kofler}}, \ and\ \bibinfo {author} {\bibfnamefont {Anton}\
  \bibnamefont {Zeilinger}},\ }\bibfield  {title} {\enquote {\bibinfo {title}
  {A snapshot of foundational attitudes toward quantum mechanics},}\ }\href
  {\doibase https://doi.org/10.1016/j.shpsb.2013.04.004} {\bibfield  {journal}
  {\bibinfo  {journal} {Studies in History and Philosophy of Science Part B:
  Studies in History and Philosophy of Modern Physics}\ }\textbf {\bibinfo
  {volume} {44}},\ \bibinfo {pages} {222 -- 230} (\bibinfo {year}
  {2013})}\BibitemShut {NoStop}%
\bibitem [{\citenamefont {Clerk}\ \emph {et~al.}(2010)\citenamefont {Clerk},
  \citenamefont {Devoret}, \citenamefont {Girvin}, \citenamefont {Marquardt},\
  and\ \citenamefont {Schoelkopf}}]{clerk2010introduction}%
  \BibitemOpen
  \bibfield  {author} {\bibinfo {author} {\bibfnamefont {A.~A.}\ \bibnamefont
  {Clerk}}, \bibinfo {author} {\bibfnamefont {M.~H.}\ \bibnamefont {Devoret}},
  \bibinfo {author} {\bibfnamefont {S.~M.}\ \bibnamefont {Girvin}}, \bibinfo
  {author} {\bibfnamefont {Florian}\ \bibnamefont {Marquardt}}, \ and\ \bibinfo
  {author} {\bibfnamefont {R.~J.}\ \bibnamefont {Schoelkopf}},\ }\bibfield
  {title} {\enquote {\bibinfo {title} {Introduction to quantum noise,
  measurement, and amplification},}\ }\href {\doibase
  10.1103/RevModPhys.82.1155} {\bibfield  {journal} {\bibinfo  {journal} {Rev.
  Mod. Phys.}\ }\textbf {\bibinfo {volume} {82}},\ \bibinfo {pages}
  {1155--1208} (\bibinfo {year} {2010})}\BibitemShut {NoStop}%
\bibitem [{\citenamefont {Campbell}\ \emph {et~al.}(2017)\citenamefont
  {Campbell}, \citenamefont {Terhal},\ and\ \citenamefont
  {Vuillot}}]{Campbell2017}%
  \BibitemOpen
  \bibfield  {author} {\bibinfo {author} {\bibfnamefont {Earl~T}\ \bibnamefont
  {Campbell}}, \bibinfo {author} {\bibfnamefont {Barbara~M}\ \bibnamefont
  {Terhal}}, \ and\ \bibinfo {author} {\bibfnamefont {Christophe}\ \bibnamefont
  {Vuillot}},\ }\bibfield  {title} {\enquote {\bibinfo {title} {{Roads towards
  fault-tolerant universal quantum computation}},}\ }\href
  {https://doi.org/10.1038/nature23460 http://10.0.4.14/nature23460
  https://www.nature.com/articles/nature23460{\#}supplementary-information}
  {\bibfield  {journal} {\bibinfo  {journal} {Nature}\ }\textbf {\bibinfo
  {volume} {549}},\ \bibinfo {pages} {172} (\bibinfo {year}
  {2017})}\BibitemShut {NoStop}%
\bibitem [{\citenamefont {Wallraff}\ \emph {et~al.}(2004)\citenamefont
  {Wallraff}, \citenamefont {Schuster}, \citenamefont {Blais}, \citenamefont
  {Frunzio}, \citenamefont {Huang}, \citenamefont {Majer}, \citenamefont
  {Kumar}, \citenamefont {Girvin},\ and\ \citenamefont
  {Schoelkopf}}]{wallraff2004strong}%
  \BibitemOpen
  \bibfield  {author} {\bibinfo {author} {\bibfnamefont {A.}~\bibnamefont
  {Wallraff}}, \bibinfo {author} {\bibfnamefont {D.~I.}\ \bibnamefont
  {Schuster}}, \bibinfo {author} {\bibfnamefont {A.}~\bibnamefont {Blais}},
  \bibinfo {author} {\bibfnamefont {L.}~\bibnamefont {Frunzio}}, \bibinfo
  {author} {\bibfnamefont {R.~S.}\ \bibnamefont {Huang}}, \bibinfo {author}
  {\bibfnamefont {J.}~\bibnamefont {Majer}}, \bibinfo {author} {\bibfnamefont
  {S.}~\bibnamefont {Kumar}}, \bibinfo {author} {\bibfnamefont {S.~M.}\
  \bibnamefont {Girvin}}, \ and\ \bibinfo {author} {\bibfnamefont {R.~J.}\
  \bibnamefont {Schoelkopf}},\ }\bibfield  {title} {\enquote {\bibinfo {title}
  {Strong coupling of a single photon to a superconducting qubit using circuit
  quantum electrodynamics},}\ }\href {\doibase 10.1038/nature02851} {\bibfield
  {journal} {\bibinfo  {journal} {Nature}\ }\textbf {\bibinfo {volume} {431}},\
  \bibinfo {pages} {162--167} (\bibinfo {year} {2004})}\BibitemShut {NoStop}%
\bibitem [{\citenamefont {Wineland}\ \emph {et~al.}(1980)\citenamefont
  {Wineland}, \citenamefont {Bergquist}, \citenamefont {Itano},\ and\
  \citenamefont {Drullinger}}]{wineland1980double}%
  \BibitemOpen
  \bibfield  {author} {\bibinfo {author} {\bibfnamefont {D.~J.}\ \bibnamefont
  {Wineland}}, \bibinfo {author} {\bibfnamefont {J.~C.}\ \bibnamefont
  {Bergquist}}, \bibinfo {author} {\bibfnamefont {Wayne~M.}\ \bibnamefont
  {Itano}}, \ and\ \bibinfo {author} {\bibfnamefont {R.~E.}\ \bibnamefont
  {Drullinger}},\ }\bibfield  {title} {\enquote {\bibinfo {title}
  {Double-resonance and optical-pumping experiments on electromagnetically
  confined, laser-cooled ions},}\ }\href {\doibase 10.1364/OL.5.000245}
  {\bibfield  {journal} {\bibinfo  {journal} {Opt. Lett.}\ }\textbf {\bibinfo
  {volume} {5}},\ \bibinfo {pages} {245--247} (\bibinfo {year}
  {1980})}\BibitemShut {NoStop}%
\bibitem [{\citenamefont {Wecker}\ \emph {et~al.}(2015)\citenamefont {Wecker},
  \citenamefont {Hastings},\ and\ \citenamefont {Troyer}}]{wecker2015progress}%
  \BibitemOpen
  \bibfield  {author} {\bibinfo {author} {\bibfnamefont {Dave}\ \bibnamefont
  {Wecker}}, \bibinfo {author} {\bibfnamefont {Matthew~B.}\ \bibnamefont
  {Hastings}}, \ and\ \bibinfo {author} {\bibfnamefont {Matthias}\ \bibnamefont
  {Troyer}},\ }\bibfield  {title} {\enquote {\bibinfo {title} {Progress towards
  practical quantum variational algorithms},}\ }\href {\doibase
  10.1103/PhysRevA.92.042303} {\bibfield  {journal} {\bibinfo  {journal} {Phys.
  Rev. A}\ }\textbf {\bibinfo {volume} {92}},\ \bibinfo {pages} {042303}
  (\bibinfo {year} {2015})}\BibitemShut {NoStop}%
\bibitem [{\citenamefont {{Jena}}\ \emph {et~al.}(2019)\citenamefont {{Jena}},
  \citenamefont {{Genin}},\ and\ \citenamefont {{Mosca}}}]{Jena2019}%
  \BibitemOpen
  \bibfield  {author} {\bibinfo {author} {\bibfnamefont {Andrew}\ \bibnamefont
  {{Jena}}}, \bibinfo {author} {\bibfnamefont {Scott}\ \bibnamefont {{Genin}}},
  \ and\ \bibinfo {author} {\bibfnamefont {Michele}\ \bibnamefont {{Mosca}}},\
  }\bibfield  {title} {\enquote {\bibinfo {title} {{Pauli Partitioning with
  Respect to Gate Sets}},}\ }\href@noop {} {\bibfield  {journal} {\bibinfo
  {journal} {arXiv e-prints}\ ,\ \bibinfo {eid} {arXiv:1907.07859}} (\bibinfo
  {year} {2019})},\ \Eprint {http://arxiv.org/abs/1907.07859} {arXiv:1907.07859
  [quant-ph]} \BibitemShut {NoStop}%
\bibitem [{\citenamefont {{Yen}}\ \emph {et~al.}(2019)\citenamefont {{Yen}},
  \citenamefont {{Verteletskyi}},\ and\ \citenamefont {{Izmaylov}}}]{Yen2019b}%
  \BibitemOpen
  \bibfield  {author} {\bibinfo {author} {\bibfnamefont {Tzu-Ching}\
  \bibnamefont {{Yen}}}, \bibinfo {author} {\bibfnamefont {Vladyslav}\
  \bibnamefont {{Verteletskyi}}}, \ and\ \bibinfo {author} {\bibfnamefont
  {Artur~F.}\ \bibnamefont {{Izmaylov}}},\ }\bibfield  {title} {\enquote
  {\bibinfo {title} {{Measuring all compatible operators in one series of a
  single-qubit measurements using unitary transformations}},}\ }\href@noop {}
  {\bibfield  {journal} {\bibinfo  {journal} {arXiv e-prints}\ ,\ \bibinfo
  {eid} {arXiv:1907.09386}} (\bibinfo {year} {2019})},\ \Eprint
  {http://arxiv.org/abs/1907.09386} {arXiv:1907.09386 [quant-ph]} \BibitemShut
  {NoStop}%
\bibitem [{\citenamefont {{Huggins}}\ \emph {et~al.}(2019)\citenamefont
  {{Huggins}}, \citenamefont {{McClean}}, \citenamefont {{Rubin}},
  \citenamefont {{Jiang}}, \citenamefont {{Wiebe}}, \citenamefont {{Whaley}},\
  and\ \citenamefont {{Babbush}}}]{Huggins2019}%
  \BibitemOpen
  \bibfield  {author} {\bibinfo {author} {\bibfnamefont {William~J.}\
  \bibnamefont {{Huggins}}}, \bibinfo {author} {\bibfnamefont {Jarrod}\
  \bibnamefont {{McClean}}}, \bibinfo {author} {\bibfnamefont {Nicholas}\
  \bibnamefont {{Rubin}}}, \bibinfo {author} {\bibfnamefont {Zhang}\
  \bibnamefont {{Jiang}}}, \bibinfo {author} {\bibfnamefont {Nathan}\
  \bibnamefont {{Wiebe}}}, \bibinfo {author} {\bibfnamefont {K.~Birgitta}\
  \bibnamefont {{Whaley}}}, \ and\ \bibinfo {author} {\bibfnamefont {Ryan}\
  \bibnamefont {{Babbush}}},\ }\bibfield  {title} {\enquote {\bibinfo {title}
  {{Efficient and Noise Resilient Measurements for Quantum Chemistry on
  Near-Term Quantum Computers}},}\ }\href@noop {} {\bibfield  {journal}
  {\bibinfo  {journal} {arXiv e-prints}\ ,\ \bibinfo {eid} {arXiv:1907.13117}}
  (\bibinfo {year} {2019})},\ \Eprint {http://arxiv.org/abs/1907.13117}
  {arXiv:1907.13117 [quant-ph]} \BibitemShut {NoStop}%
\bibitem [{\citenamefont {{Gokhale}}\ \emph {et~al.}(2019)\citenamefont
  {{Gokhale}}, \citenamefont {{Angiuli}}, \citenamefont {{Ding}}, \citenamefont
  {{Gui}}, \citenamefont {{Tomesh}}, \citenamefont {{Suchara}}, \citenamefont
  {{Martonosi}},\ and\ \citenamefont {{Chong}}}]{Gokhale2019}%
  \BibitemOpen
  \bibfield  {author} {\bibinfo {author} {\bibfnamefont {Pranav}\ \bibnamefont
  {{Gokhale}}}, \bibinfo {author} {\bibfnamefont {Olivia}\ \bibnamefont
  {{Angiuli}}}, \bibinfo {author} {\bibfnamefont {Yongshan}\ \bibnamefont
  {{Ding}}}, \bibinfo {author} {\bibfnamefont {Kaiwen}\ \bibnamefont {{Gui}}},
  \bibinfo {author} {\bibfnamefont {Teague}\ \bibnamefont {{Tomesh}}}, \bibinfo
  {author} {\bibfnamefont {Martin}\ \bibnamefont {{Suchara}}}, \bibinfo
  {author} {\bibfnamefont {Margaret}\ \bibnamefont {{Martonosi}}}, \ and\
  \bibinfo {author} {\bibfnamefont {Frederic~T.}\ \bibnamefont {{Chong}}},\
  }\bibfield  {title} {\enquote {\bibinfo {title} {{Minimizing State
  Preparations in Variational Quantum Eigensolver by Partitioning into
  Commuting Families}},}\ }\href@noop {} {\bibfield  {journal} {\bibinfo
  {journal} {arXiv e-prints}\ ,\ \bibinfo {eid} {arXiv:1907.13623}} (\bibinfo
  {year} {2019})},\ \Eprint {http://arxiv.org/abs/1907.13623} {arXiv:1907.13623
  [quant-ph]} \BibitemShut {NoStop}%
\bibitem [{\citenamefont {{Crawford}}\ \emph {et~al.}(2019)\citenamefont
  {{Crawford}}, \citenamefont {{van Straaten}}, \citenamefont {{Wang}},
  \citenamefont {{Parks}}, \citenamefont {{Campbell}},\ and\ \citenamefont
  {{Brierley}}}]{Crawford2019}%
  \BibitemOpen
  \bibfield  {author} {\bibinfo {author} {\bibfnamefont {Ophelia}\ \bibnamefont
  {{Crawford}}}, \bibinfo {author} {\bibfnamefont {Barnaby}\ \bibnamefont {{van
  Straaten}}}, \bibinfo {author} {\bibfnamefont {Daochen}\ \bibnamefont
  {{Wang}}}, \bibinfo {author} {\bibfnamefont {Thomas}\ \bibnamefont
  {{Parks}}}, \bibinfo {author} {\bibfnamefont {Earl}\ \bibnamefont
  {{Campbell}}}, \ and\ \bibinfo {author} {\bibfnamefont {Stephen}\
  \bibnamefont {{Brierley}}},\ }\bibfield  {title} {\enquote {\bibinfo {title}
  {{Efficient quantum measurement of Pauli operators}},}\ }\href@noop {}
  {\bibfield  {journal} {\bibinfo  {journal} {arXiv e-prints}\ ,\ \bibinfo
  {eid} {arXiv:1908.06942}} (\bibinfo {year} {2019})},\ \Eprint
  {http://arxiv.org/abs/1908.06942} {arXiv:1908.06942 [quant-ph]} \BibitemShut
  {NoStop}%
\bibitem [{\citenamefont {{Zhao}}\ \emph {et~al.}(2019)\citenamefont {{Zhao}},
  \citenamefont {{Tranter}}, \citenamefont {{Kirby}}, \citenamefont {{Ung}},
  \citenamefont {{Miyake}},\ and\ \citenamefont
  {{Love}}}]{zhao2019measurement}%
  \BibitemOpen
  \bibfield  {author} {\bibinfo {author} {\bibfnamefont {Andrew}\ \bibnamefont
  {{Zhao}}}, \bibinfo {author} {\bibfnamefont {Andrew}\ \bibnamefont
  {{Tranter}}}, \bibinfo {author} {\bibfnamefont {William~M.}\ \bibnamefont
  {{Kirby}}}, \bibinfo {author} {\bibfnamefont {Shu~Fay}\ \bibnamefont
  {{Ung}}}, \bibinfo {author} {\bibfnamefont {Akimasa}\ \bibnamefont
  {{Miyake}}}, \ and\ \bibinfo {author} {\bibfnamefont {Peter}\ \bibnamefont
  {{Love}}},\ }\bibfield  {title} {\enquote {\bibinfo {title} {{Measurement
  reduction in variational quantum algorithms}},}\ }\href@noop {} {\bibfield
  {journal} {\bibinfo  {journal} {arXiv e-prints}\ ,\ \bibinfo {eid}
  {arXiv:1908.08067}} (\bibinfo {year} {2019})},\ \Eprint
  {http://arxiv.org/abs/1908.08067} {arXiv:1908.08067 [quant-ph]} \BibitemShut
  {NoStop}%
\bibitem [{\citenamefont {{Carleo}}\ \emph {et~al.}(2019)\citenamefont
  {{Carleo}}, \citenamefont {{Cirac}}, \citenamefont {{Cranmer}}, \citenamefont
  {{Daudet}}, \citenamefont {{Schuld}}, \citenamefont {{Tishby}}, \citenamefont
  {{Vogt-Maranto}},\ and\ \citenamefont {{Zdeborov{\'a}}}}]{MLQreview}%
  \BibitemOpen
  \bibfield  {author} {\bibinfo {author} {\bibfnamefont {Giuseppe}\
  \bibnamefont {{Carleo}}}, \bibinfo {author} {\bibfnamefont {Ignacio}\
  \bibnamefont {{Cirac}}}, \bibinfo {author} {\bibfnamefont {Kyle}\
  \bibnamefont {{Cranmer}}}, \bibinfo {author} {\bibfnamefont {Laurent}\
  \bibnamefont {{Daudet}}}, \bibinfo {author} {\bibfnamefont {Maria}\
  \bibnamefont {{Schuld}}}, \bibinfo {author} {\bibfnamefont {Naftali}\
  \bibnamefont {{Tishby}}}, \bibinfo {author} {\bibfnamefont {Leslie}\
  \bibnamefont {{Vogt-Maranto}}}, \ and\ \bibinfo {author} {\bibfnamefont
  {Lenka}\ \bibnamefont {{Zdeborov{\'a}}}},\ }\bibfield  {title} {\enquote
  {\bibinfo {title} {{Machine learning and the physical sciences}},}\
  }\href@noop {} {\bibfield  {journal} {\bibinfo  {journal} {arXiv e-prints}\ }
  (\bibinfo {year} {2019})},\ \Eprint {http://arxiv.org/abs/1903.10563}
  {arXiv:1903.10563} \BibitemShut {NoStop}%
\bibitem [{\citenamefont {Carrasquilla}\ and\ \citenamefont
  {Melko}(2017)}]{Carrasquilla17}%
  \BibitemOpen
  \bibfield  {author} {\bibinfo {author} {\bibfnamefont {Juan}\ \bibnamefont
  {Carrasquilla}}\ and\ \bibinfo {author} {\bibfnamefont {Roger~G.}\
  \bibnamefont {Melko}},\ }\bibfield  {title} {\enquote {\bibinfo {title}
  {Machine learning phases of matter},}\ }\href
  {https://doi.org/10.1038/nphys4035} {\bibfield  {journal} {\bibinfo
  {journal} {Nature Physics}\ }\textbf {\bibinfo {volume} {13}},\ \bibinfo
  {pages} {431} (\bibinfo {year} {2017})}\BibitemShut {NoStop}%
\bibitem [{\citenamefont {Wang}(2016)}]{LeiWang16}%
  \BibitemOpen
  \bibfield  {author} {\bibinfo {author} {\bibfnamefont {Lei}\ \bibnamefont
  {Wang}},\ }\bibfield  {title} {\enquote {\bibinfo {title} {Discovering phase
  transitions with unsupervised learning},}\ }\href {\doibase
  10.1103/PhysRevB.94.195105} {\bibfield  {journal} {\bibinfo  {journal} {Phys.
  Rev. B}\ }\textbf {\bibinfo {volume} {94}},\ \bibinfo {pages} {195105}
  (\bibinfo {year} {2016})}\BibitemShut {NoStop}%
\bibitem [{\citenamefont {Carleo}\ and\ \citenamefont
  {Troyer}(2017)}]{Carleo17}%
  \BibitemOpen
  \bibfield  {author} {\bibinfo {author} {\bibfnamefont {Giuseppe}\
  \bibnamefont {Carleo}}\ and\ \bibinfo {author} {\bibfnamefont {Matthias}\
  \bibnamefont {Troyer}},\ }\bibfield  {title} {\enquote {\bibinfo {title}
  {Solving the quantum many-body problem with artificial neural networks},}\
  }\href {\doibase 10.1126/science.aag2302} {\bibfield  {journal} {\bibinfo
  {journal} {Science}\ }\textbf {\bibinfo {volume} {355}},\ \bibinfo {pages}
  {602--606} (\bibinfo {year} {2017})}\BibitemShut {NoStop}%
\bibitem [{\citenamefont {Torlai}\ and\ \citenamefont
  {Melko}(2016)}]{Torlai16}%
  \BibitemOpen
  \bibfield  {author} {\bibinfo {author} {\bibfnamefont {Giacomo}\ \bibnamefont
  {Torlai}}\ and\ \bibinfo {author} {\bibfnamefont {Roger~G}\ \bibnamefont
  {Melko}},\ }\bibfield  {title} {\enquote {\bibinfo {title} {{Learning
  thermodynamics with Boltzmann machines}},}\ }\href {\doibase
  10.1103/PhysRevB.94.165134} {\bibfield  {journal} {\bibinfo  {journal}
  {Physical Review B}\ }\textbf {\bibinfo {volume} {94}},\ \bibinfo {pages}
  {165134} (\bibinfo {year} {2016})}\BibitemShut {NoStop}%
\bibitem [{\citenamefont {van Nieuwenburg}\ \emph {et~al.}(2017)\citenamefont
  {van Nieuwenburg}, \citenamefont {Liu},\ and\ \citenamefont
  {Huber}}]{Evert17}%
  \BibitemOpen
  \bibfield  {author} {\bibinfo {author} {\bibfnamefont {Evert P.~L.}\
  \bibnamefont {van Nieuwenburg}}, \bibinfo {author} {\bibfnamefont {Ye-Hua}\
  \bibnamefont {Liu}}, \ and\ \bibinfo {author} {\bibfnamefont {Sebastian~D.}\
  \bibnamefont {Huber}},\ }\bibfield  {title} {\enquote {\bibinfo {title}
  {Learning phase transitions by confusion},}\ }\href
  {https://doi.org/10.1038/nphys4037} {\bibfield  {journal} {\bibinfo
  {journal} {Nature Physics}\ }\textbf {\bibinfo {volume} {13}},\ \bibinfo
  {pages} {435} (\bibinfo {year} {2017})}\BibitemShut {NoStop}%
\bibitem [{\citenamefont {Koch-Janusz}\ and\ \citenamefont
  {Ringel}(2018)}]{neural_rg}%
  \BibitemOpen
  \bibfield  {author} {\bibinfo {author} {\bibfnamefont {Maciej}\ \bibnamefont
  {Koch-Janusz}}\ and\ \bibinfo {author} {\bibfnamefont {Zohar}\ \bibnamefont
  {Ringel}},\ }\bibfield  {title} {\enquote {\bibinfo {title} {Mutual
  information, neural networks and the renormalization group},}\ }\href
  {\doibase 10.1038/s41567-018-0081-4} {\bibfield  {journal} {\bibinfo
  {journal} {Nature Physics}\ }\textbf {\bibinfo {volume} {14}},\ \bibinfo
  {pages} {578--582} (\bibinfo {year} {2018})}\BibitemShut {NoStop}%
\bibitem [{\citenamefont {Torlai}\ \emph {et~al.}(2018)\citenamefont {Torlai},
  \citenamefont {Mazzola}, \citenamefont {Carrasquilla}, \citenamefont
  {Troyer}, \citenamefont {Melko},\ and\ \citenamefont
  {Carleo}}]{torlai_2018_nnqst}%
  \BibitemOpen
  \bibfield  {author} {\bibinfo {author} {\bibfnamefont {Giacomo}\ \bibnamefont
  {Torlai}}, \bibinfo {author} {\bibfnamefont {Guglielmo}\ \bibnamefont
  {Mazzola}}, \bibinfo {author} {\bibfnamefont {Juan}\ \bibnamefont
  {Carrasquilla}}, \bibinfo {author} {\bibfnamefont {Matthias}\ \bibnamefont
  {Troyer}}, \bibinfo {author} {\bibfnamefont {Roger}\ \bibnamefont {Melko}}, \
  and\ \bibinfo {author} {\bibfnamefont {Giuseppe}\ \bibnamefont {Carleo}},\
  }\bibfield  {title} {\enquote {\bibinfo {title} {{Neural-network quantum
  state tomography}},}\ }\href {\doibase 10.1038/s41567-018-0048-5} {\bibfield
  {journal} {\bibinfo  {journal} {Nature Physics}\ }\textbf {\bibinfo {volume}
  {14}},\ \bibinfo {pages} {447--450} (\bibinfo {year} {2018})}\BibitemShut
  {NoStop}%
\bibitem [{\citenamefont {Bukov}\ \emph {et~al.}(2018)\citenamefont {Bukov},
  \citenamefont {Day}, \citenamefont {Sels}, \citenamefont {Weinberg},
  \citenamefont {Polkovnikov},\ and\ \citenamefont {Mehta}}]{bukov18}%
  \BibitemOpen
  \bibfield  {author} {\bibinfo {author} {\bibfnamefont {Marin}\ \bibnamefont
  {Bukov}}, \bibinfo {author} {\bibfnamefont {Alexandre G.~R.}\ \bibnamefont
  {Day}}, \bibinfo {author} {\bibfnamefont {Dries}\ \bibnamefont {Sels}},
  \bibinfo {author} {\bibfnamefont {Phillip}\ \bibnamefont {Weinberg}},
  \bibinfo {author} {\bibfnamefont {Anatoli}\ \bibnamefont {Polkovnikov}}, \
  and\ \bibinfo {author} {\bibfnamefont {Pankaj}\ \bibnamefont {Mehta}},\
  }\bibfield  {title} {\enquote {\bibinfo {title} {Reinforcement learning in
  different phases of quantum control},}\ }\href {\doibase
  10.1103/PhysRevX.8.031086} {\bibfield  {journal} {\bibinfo  {journal} {Phys.
  Rev. X}\ }\textbf {\bibinfo {volume} {8}},\ \bibinfo {pages} {031086}
  (\bibinfo {year} {2018})}\BibitemShut {NoStop}%
\bibitem [{\citenamefont {Seif}\ \emph {et~al.}(2018)\citenamefont {Seif},
  \citenamefont {Landsman}, \citenamefont {Linke}, \citenamefont {Figgatt},
  \citenamefont {Monroe},\ and\ \citenamefont {Hafezi}}]{Seif_2018}%
  \BibitemOpen
  \bibfield  {author} {\bibinfo {author} {\bibfnamefont {Alireza}\ \bibnamefont
  {Seif}}, \bibinfo {author} {\bibfnamefont {Kevin~A}\ \bibnamefont
  {Landsman}}, \bibinfo {author} {\bibfnamefont {Norbert~M}\ \bibnamefont
  {Linke}}, \bibinfo {author} {\bibfnamefont {Caroline}\ \bibnamefont
  {Figgatt}}, \bibinfo {author} {\bibfnamefont {C}~\bibnamefont {Monroe}}, \
  and\ \bibinfo {author} {\bibfnamefont {Mohammad}\ \bibnamefont {Hafezi}},\
  }\bibfield  {title} {\enquote {\bibinfo {title} {Machine learning assisted
  readout of trapped-ion qubits},}\ }\href {\doibase 10.1088/1361-6455/aad62b}
  {\bibfield  {journal} {\bibinfo  {journal} {Journal of Physics B: Atomic,
  Molecular and Optical Physics}\ }\textbf {\bibinfo {volume} {51}},\ \bibinfo
  {pages} {174006} (\bibinfo {year} {2018})}\BibitemShut {NoStop}%
\bibitem [{\citenamefont {Rem}\ \emph {et~al.}(2019)\citenamefont {Rem},
  \citenamefont {K{\"a}ming}, \citenamefont {Tarnowski}, \citenamefont
  {Asteria}, \citenamefont {Fl{\"a}schner}, \citenamefont {Becker},
  \citenamefont {Sengstock},\ and\ \citenamefont
  {Weitenberg}}]{Rem_MLcoldatoms}%
  \BibitemOpen
  \bibfield  {author} {\bibinfo {author} {\bibfnamefont {Benno~S.}\
  \bibnamefont {Rem}}, \bibinfo {author} {\bibfnamefont {Niklas}\ \bibnamefont
  {K{\"a}ming}}, \bibinfo {author} {\bibfnamefont {Matthias}\ \bibnamefont
  {Tarnowski}}, \bibinfo {author} {\bibfnamefont {Luca}\ \bibnamefont
  {Asteria}}, \bibinfo {author} {\bibfnamefont {Nick}\ \bibnamefont
  {Fl{\"a}schner}}, \bibinfo {author} {\bibfnamefont {Christoph}\ \bibnamefont
  {Becker}}, \bibinfo {author} {\bibfnamefont {Klaus}\ \bibnamefont
  {Sengstock}}, \ and\ \bibinfo {author} {\bibfnamefont {Christof}\
  \bibnamefont {Weitenberg}},\ }\bibfield  {title} {\enquote {\bibinfo {title}
  {Identifying quantum phase transitions using artificial neural networks on
  experimental data},}\ }\href {https://doi.org/10.1038/s41567-019-0554-0}
  {\bibfield  {journal} {\bibinfo  {journal} {Nature Physics}\ } (\bibinfo
  {year} {2019})}\BibitemShut {NoStop}%
\bibitem [{\citenamefont {Bohrdt}\ \emph {et~al.}(2019)\citenamefont {Bohrdt},
  \citenamefont {Chiu}, \citenamefont {Ji}, \citenamefont {Xu}, \citenamefont
  {Greif}, \citenamefont {Greiner}, \citenamefont {Demler}, \citenamefont
  {Grusdt},\ and\ \citenamefont {Knap}}]{Bohrdt_MLfermions}%
  \BibitemOpen
  \bibfield  {author} {\bibinfo {author} {\bibfnamefont {Annabelle}\
  \bibnamefont {Bohrdt}}, \bibinfo {author} {\bibfnamefont {Christie~S.}\
  \bibnamefont {Chiu}}, \bibinfo {author} {\bibfnamefont {Geoffrey}\
  \bibnamefont {Ji}}, \bibinfo {author} {\bibfnamefont {Muqing}\ \bibnamefont
  {Xu}}, \bibinfo {author} {\bibfnamefont {Daniel}\ \bibnamefont {Greif}},
  \bibinfo {author} {\bibfnamefont {Markus}\ \bibnamefont {Greiner}}, \bibinfo
  {author} {\bibfnamefont {Eugene}\ \bibnamefont {Demler}}, \bibinfo {author}
  {\bibfnamefont {Fabian}\ \bibnamefont {Grusdt}}, \ and\ \bibinfo {author}
  {\bibfnamefont {Michael}\ \bibnamefont {Knap}},\ }\bibfield  {title}
  {\enquote {\bibinfo {title} {Classifying snapshots of the doped hubbard model
  with machine learning},}\ }\href {https://doi.org/10.1038/s41567-019-0565-x}
  {\bibfield  {journal} {\bibinfo  {journal} {Nature Physics}\ } (\bibinfo
  {year} {2019})}\BibitemShut {NoStop}%
\bibitem [{\citenamefont {{Torlai}}\ \emph {et~al.}(2019)\citenamefont
  {{Torlai}}, \citenamefont {{Timar}}, \citenamefont {{van Nieuwenburg}},
  \citenamefont {{Levine}}, \citenamefont {{Omran}}, \citenamefont
  {{Keesling}}, \citenamefont {{Bernien}}, \citenamefont {{Greiner}},
  \citenamefont {{Vuleti{\'c}}}, \citenamefont {{Lukin}}, \citenamefont
  {{Melko}},\ and\ \citenamefont {{Endres}}}]{torlai_rydberg}%
  \BibitemOpen
  \bibfield  {author} {\bibinfo {author} {\bibfnamefont {Giacomo}\ \bibnamefont
  {{Torlai}}}, \bibinfo {author} {\bibfnamefont {Brian}\ \bibnamefont
  {{Timar}}}, \bibinfo {author} {\bibfnamefont {Evert P.~L.}\ \bibnamefont
  {{van Nieuwenburg}}}, \bibinfo {author} {\bibfnamefont {Harry}\ \bibnamefont
  {{Levine}}}, \bibinfo {author} {\bibfnamefont {Ahmed}\ \bibnamefont
  {{Omran}}}, \bibinfo {author} {\bibfnamefont {Alexander}\ \bibnamefont
  {{Keesling}}}, \bibinfo {author} {\bibfnamefont {Hannes}\ \bibnamefont
  {{Bernien}}}, \bibinfo {author} {\bibfnamefont {Markus}\ \bibnamefont
  {{Greiner}}}, \bibinfo {author} {\bibfnamefont {Vladan}\ \bibnamefont
  {{Vuleti{\'c}}}}, \bibinfo {author} {\bibfnamefont {Mikhail~D.}\ \bibnamefont
  {{Lukin}}}, \bibinfo {author} {\bibfnamefont {Roger~G.}\ \bibnamefont
  {{Melko}}}, \ and\ \bibinfo {author} {\bibfnamefont {Manuel}\ \bibnamefont
  {{Endres}}},\ }\bibfield  {title} {\enquote {\bibinfo {title} {{Integrating
  Neural Networks with a Quantum Simulator for State Reconstruction}},}\
  }\href@noop {} {\bibfield  {journal} {\bibinfo  {journal} {arXiv e-prints}\
  ,\ \bibinfo {eid} {arXiv:1904.08441}} (\bibinfo {year} {2019})},\ \Eprint
  {http://arxiv.org/abs/1904.08441} {arXiv:1904.08441 [quant-ph]} \BibitemShut
  {NoStop}%
\bibitem [{\citenamefont {Zhang}\ \emph {et~al.}(2019)\citenamefont {Zhang},
  \citenamefont {Mesaros}, \citenamefont {Fujita}, \citenamefont {Edkins},
  \citenamefont {Hamidian}, \citenamefont {Ch'ng}, \citenamefont {Eisaki},
  \citenamefont {Uchida}, \citenamefont {Davis}, \citenamefont {Khatami},\ and\
  \citenamefont {Kim}}]{Zhang_MLcuprates}%
  \BibitemOpen
  \bibfield  {author} {\bibinfo {author} {\bibfnamefont {Yi}~\bibnamefont
  {Zhang}}, \bibinfo {author} {\bibfnamefont {A.}~\bibnamefont {Mesaros}},
  \bibinfo {author} {\bibfnamefont {K.}~\bibnamefont {Fujita}}, \bibinfo
  {author} {\bibfnamefont {S.~D.}\ \bibnamefont {Edkins}}, \bibinfo {author}
  {\bibfnamefont {M.~H.}\ \bibnamefont {Hamidian}}, \bibinfo {author}
  {\bibfnamefont {K.}~\bibnamefont {Ch'ng}}, \bibinfo {author} {\bibfnamefont
  {H.}~\bibnamefont {Eisaki}}, \bibinfo {author} {\bibfnamefont
  {S.}~\bibnamefont {Uchida}}, \bibinfo {author} {\bibfnamefont
  {J.~C.~S{\'e}amus}\ \bibnamefont {Davis}}, \bibinfo {author} {\bibfnamefont
  {Ehsan}\ \bibnamefont {Khatami}}, \ and\ \bibinfo {author} {\bibfnamefont
  {Eun-Ah}\ \bibnamefont {Kim}},\ }\bibfield  {title} {\enquote {\bibinfo
  {title} {Machine learning in electronic-quantum-matter imaging
  experiments},}\ }\href {\doibase 10.1038/s41586-019-1319-8} {\bibfield
  {journal} {\bibinfo  {journal} {Nature}\ }\textbf {\bibinfo {volume} {570}},\
  \bibinfo {pages} {484--490} (\bibinfo {year} {2019})}\BibitemShut {NoStop}%
\bibitem [{\citenamefont {{Teoh}}\ \emph {et~al.}(2019)\citenamefont {{Teoh}},
  \citenamefont {{Drygala}}, \citenamefont {{Melko}},\ and\ \citenamefont
  {{Islam}}}]{islam2019}%
  \BibitemOpen
  \bibfield  {author} {\bibinfo {author} {\bibfnamefont {Yi~Hong}\ \bibnamefont
  {{Teoh}}}, \bibinfo {author} {\bibfnamefont {Marina}\ \bibnamefont
  {{Drygala}}}, \bibinfo {author} {\bibfnamefont {Roger~G.}\ \bibnamefont
  {{Melko}}}, \ and\ \bibinfo {author} {\bibfnamefont {Rajibul}\ \bibnamefont
  {{Islam}}},\ }\bibfield  {title} {\enquote {\bibinfo {title} {{Machine
  learning design of a trapped-ion quantum spin simulator}},}\ }\href@noop {}
  {\bibfield  {journal} {\bibinfo  {journal} {arXiv e-prints}\ ,\ \bibinfo
  {eid} {arXiv:1910.02496}} (\bibinfo {year} {2019})},\ \Eprint
  {http://arxiv.org/abs/1910.02496} {arXiv:1910.02496 [quant-ph]} \BibitemShut
  {NoStop}%
\bibitem [{\citenamefont {Preskill}(2018)}]{Preskill2018}%
  \BibitemOpen
  \bibfield  {author} {\bibinfo {author} {\bibfnamefont {John}\ \bibnamefont
  {Preskill}},\ }\bibfield  {title} {\enquote {\bibinfo {title} {Quantum
  {C}omputing in the {NISQ} era and beyond},}\ }\href {\doibase
  10.22331/q-2018-08-06-79} {\bibfield  {journal} {\bibinfo  {journal}
  {{Quantum}}\ }\textbf {\bibinfo {volume} {2}},\ \bibinfo {pages} {79}
  (\bibinfo {year} {2018})}\BibitemShut {NoStop}%
\bibitem [{\citenamefont {Banaszek}\ \emph {et~al.}(2013)\citenamefont
  {Banaszek}, \citenamefont {Cramer},\ and\ \citenamefont
  {Gross}}]{Banaszek2013}%
  \BibitemOpen
  \bibfield  {author} {\bibinfo {author} {\bibfnamefont {K}~\bibnamefont
  {Banaszek}}, \bibinfo {author} {\bibfnamefont {M}~\bibnamefont {Cramer}}, \
  and\ \bibinfo {author} {\bibfnamefont {D}~\bibnamefont {Gross}},\ }\bibfield
  {title} {\enquote {\bibinfo {title} {{Focus on quantum tomography}},}\ }\href
  {\doibase 10.1088/1367-2630/15/12/125020} {\bibfield  {journal} {\bibinfo
  {journal} {New Journal of Physics}\ }\textbf {\bibinfo {volume} {15}},\
  \bibinfo {pages} {125020} (\bibinfo {year} {2013})}\BibitemShut {NoStop}%
\bibitem [{\citenamefont {Ackley}\ \emph {et~al.}(1985)\citenamefont {Ackley},
  \citenamefont {Hinton},\ and\ \citenamefont {Sejnowski}}]{Hinton85}%
  \BibitemOpen
  \bibfield  {author} {\bibinfo {author} {\bibfnamefont {David~H.}\
  \bibnamefont {Ackley}}, \bibinfo {author} {\bibfnamefont {Geoffrey~E.}\
  \bibnamefont {Hinton}}, \ and\ \bibinfo {author} {\bibfnamefont
  {Terrence~J.}\ \bibnamefont {Sejnowski}},\ }\bibfield  {title} {\enquote
  {\bibinfo {title} {A learning algorithm for boltzmann machines*},}\ }\href
  {\doibase 10.1207/s15516709cog0901{\_}7} {\bibfield  {journal} {\bibinfo
  {journal} {Cognitive Science}\ }\textbf {\bibinfo {volume} {9}},\ \bibinfo
  {pages} {147--169} (\bibinfo {year} {1985})}\BibitemShut {NoStop}%
\bibitem [{\citenamefont {{Torlai}}\ and\ \citenamefont
  {{Melko}}(2019)}]{mlNISQ}%
  \BibitemOpen
  \bibfield  {author} {\bibinfo {author} {\bibfnamefont {Giacomo}\ \bibnamefont
  {{Torlai}}}\ and\ \bibinfo {author} {\bibfnamefont {Roger~G.}\ \bibnamefont
  {{Melko}}},\ }\bibfield  {title} {\enquote {\bibinfo {title} {{Machine
  learning quantum states in the NISQ era}},}\ }\href@noop {} {\bibfield
  {journal} {\bibinfo  {journal} {arXiv e-prints}\ ,\ \bibinfo {eid}
  {arXiv:1905.04312}} (\bibinfo {year} {2019})},\ \Eprint
  {http://arxiv.org/abs/1905.04312} {arXiv:1905.04312 [quant-ph]} \BibitemShut
  {NoStop}%
\bibitem [{\citenamefont {Melko}\ \emph {et~al.}(2019)\citenamefont {Melko},
  \citenamefont {Carleo}, \citenamefont {Carrasquilla},\ and\ \citenamefont
  {Cirac}}]{RBM_natphys}%
  \BibitemOpen
  \bibfield  {author} {\bibinfo {author} {\bibfnamefont {Roger~G.}\
  \bibnamefont {Melko}}, \bibinfo {author} {\bibfnamefont {Giuseppe}\
  \bibnamefont {Carleo}}, \bibinfo {author} {\bibfnamefont {Juan}\ \bibnamefont
  {Carrasquilla}}, \ and\ \bibinfo {author} {\bibfnamefont {J.~Ignacio}\
  \bibnamefont {Cirac}},\ }\bibfield  {title} {\enquote {\bibinfo {title}
  {Restricted boltzmann machines in quantum physics},}\ }\href {\doibase
  10.1038/s41567-019-0545-1} {\bibfield  {journal} {\bibinfo  {journal} {Nature
  Physics}\ } (\bibinfo {year} {2019}),\ 10.1038/s41567-019-0545-1}\BibitemShut
  {NoStop}%
\bibitem [{\citenamefont {Poulin}\ \emph {et~al.}(2015)\citenamefont {Poulin},
  \citenamefont {Hastings}, \citenamefont {Wecker}, \citenamefont {Wiebe},
  \citenamefont {Doberty},\ and\ \citenamefont
  {Troyer}}]{DBLP:journals/qic/PoulinHWWDT15}%
  \BibitemOpen
  \bibfield  {author} {\bibinfo {author} {\bibfnamefont {David}\ \bibnamefont
  {Poulin}}, \bibinfo {author} {\bibfnamefont {Matthew~B.}\ \bibnamefont
  {Hastings}}, \bibinfo {author} {\bibfnamefont {Dave}\ \bibnamefont {Wecker}},
  \bibinfo {author} {\bibfnamefont {Nathan}\ \bibnamefont {Wiebe}}, \bibinfo
  {author} {\bibfnamefont {Andrew~C.}\ \bibnamefont {Doberty}}, \ and\ \bibinfo
  {author} {\bibfnamefont {Matthias}\ \bibnamefont {Troyer}},\ }\bibfield
  {title} {\enquote {\bibinfo {title} {The trotter step size required for
  accurate quantum simulation of quantum chemistry},}\ }\href
  {http://www.rintonpress.com/xxqic15/qic-15-56/0361-0384.pdf} {\bibfield
  {journal} {\bibinfo  {journal} {Quantum Information {\&} Computation}\
  }\textbf {\bibinfo {volume} {15}},\ \bibinfo {pages} {361--384} (\bibinfo
  {year} {2015})}\BibitemShut {NoStop}%
\bibitem [{\citenamefont {Kandala}\ \emph {et~al.}(2017)\citenamefont
  {Kandala}, \citenamefont {Mezzacapo}, \citenamefont {Temme}, \citenamefont
  {Takita}, \citenamefont {Brink}, \citenamefont {Chow},\ and\ \citenamefont
  {Gambetta}}]{Kandala17}%
  \BibitemOpen
  \bibfield  {author} {\bibinfo {author} {\bibfnamefont {Abhinav}\ \bibnamefont
  {Kandala}}, \bibinfo {author} {\bibfnamefont {Antonio}\ \bibnamefont
  {Mezzacapo}}, \bibinfo {author} {\bibfnamefont {Kristan}\ \bibnamefont
  {Temme}}, \bibinfo {author} {\bibfnamefont {Maika}\ \bibnamefont {Takita}},
  \bibinfo {author} {\bibfnamefont {Markus}\ \bibnamefont {Brink}}, \bibinfo
  {author} {\bibfnamefont {Jerry~M.}\ \bibnamefont {Chow}}, \ and\ \bibinfo
  {author} {\bibfnamefont {Jay~M.}\ \bibnamefont {Gambetta}},\ }\bibfield
  {title} {\enquote {\bibinfo {title} {Hardware-efficient variational quantum
  eigensolver for small molecules and quantum magnets},}\ }\href
  {https://doi.org/10.1038/nature23879} {\bibfield  {journal} {\bibinfo
  {journal} {Nature}\ }\textbf {\bibinfo {volume} {549}},\ \bibinfo {pages}
  {242 EP --} (\bibinfo {year} {2017})}\BibitemShut {NoStop}%
\bibitem [{\citenamefont {{Bravyi}}\ \emph {et~al.}(2017)\citenamefont
  {{Bravyi}}, \citenamefont {{Gambetta}}, \citenamefont {{Mezzacapo}},\ and\
  \citenamefont {{Temme}}}]{Bravyi17}%
  \BibitemOpen
  \bibfield  {author} {\bibinfo {author} {\bibfnamefont {Sergey}\ \bibnamefont
  {{Bravyi}}}, \bibinfo {author} {\bibfnamefont {Jay~M.}\ \bibnamefont
  {{Gambetta}}}, \bibinfo {author} {\bibfnamefont {Antonio}\ \bibnamefont
  {{Mezzacapo}}}, \ and\ \bibinfo {author} {\bibfnamefont {Kristan}\
  \bibnamefont {{Temme}}},\ }\bibfield  {title} {\enquote {\bibinfo {title}
  {{Tapering off qubits to simulate fermionic Hamiltonians}},}\ }\href@noop {}
  {\bibfield  {journal} {\bibinfo  {journal} {arXiv e-prints}\ ,\ \bibinfo
  {eid} {arXiv:1701.08213}} (\bibinfo {year} {2017})},\ \Eprint
  {http://arxiv.org/abs/1701.08213} {arXiv:1701.08213 [quant-ph]} \BibitemShut
  {NoStop}%
\bibitem [{\citenamefont {Choo}\ \emph {et~al.}(2019)\citenamefont {Choo},
  \citenamefont {Mezzacapo},\ and\ \citenamefont
  {Carleo}}]{choo_fermionic_2019}%
  \BibitemOpen
  \bibfield  {author} {\bibinfo {author} {\bibfnamefont {Kenny}\ \bibnamefont
  {Choo}}, \bibinfo {author} {\bibfnamefont {Antonio}\ \bibnamefont
  {Mezzacapo}}, \ and\ \bibinfo {author} {\bibfnamefont {Giuseppe}\
  \bibnamefont {Carleo}},\ }\bibfield  {title} {\enquote {\bibinfo {title}
  {Fermionic neural-network states for ab-initio electronic structure},}\
  }\href {http://arxiv.org/abs/1909.12852} {\bibfield  {journal} {\bibinfo
  {journal} {arXiv:1909.12852}\ } (\bibinfo {year} {2019})}\BibitemShut
  {NoStop}%
\bibitem [{\citenamefont {Kandala}\ \emph {et~al.}(2019)\citenamefont
  {Kandala}, \citenamefont {Temme}, \citenamefont {C{\'o}rcoles}, \citenamefont
  {Mezzacapo}, \citenamefont {Chow},\ and\ \citenamefont
  {Gambetta}}]{Kandala19}%
  \BibitemOpen
  \bibfield  {author} {\bibinfo {author} {\bibfnamefont {Abhinav}\ \bibnamefont
  {Kandala}}, \bibinfo {author} {\bibfnamefont {Kristan}\ \bibnamefont
  {Temme}}, \bibinfo {author} {\bibfnamefont {Antonio~D.}\ \bibnamefont
  {C{\'o}rcoles}}, \bibinfo {author} {\bibfnamefont {Antonio}\ \bibnamefont
  {Mezzacapo}}, \bibinfo {author} {\bibfnamefont {Jerry~M.}\ \bibnamefont
  {Chow}}, \ and\ \bibinfo {author} {\bibfnamefont {Jay~M.}\ \bibnamefont
  {Gambetta}},\ }\bibfield  {title} {\enquote {\bibinfo {title} {Error
  mitigation extends the computational reach of a noisy quantum processor},}\
  }\href {\doibase 10.1038/s41586-019-1040-7} {\bibfield  {journal} {\bibinfo
  {journal} {Nature}\ }\textbf {\bibinfo {volume} {567}},\ \bibinfo {pages}
  {491--495} (\bibinfo {year} {2019})}\BibitemShut {NoStop}%
\bibitem [{\citenamefont {{Sehayek}}\ \emph {et~al.}(2019)\citenamefont
  {{Sehayek}}, \citenamefont {{Golubeva}}, \citenamefont {{Albergo}},
  \citenamefont {{Kulchytskyy}}, \citenamefont {{Torlai}},\ and\ \citenamefont
  {{Melko}}}]{Sehayek2019}%
  \BibitemOpen
  \bibfield  {author} {\bibinfo {author} {\bibfnamefont {Dan}\ \bibnamefont
  {{Sehayek}}}, \bibinfo {author} {\bibfnamefont {Anna}\ \bibnamefont
  {{Golubeva}}}, \bibinfo {author} {\bibfnamefont {Michael~S.}\ \bibnamefont
  {{Albergo}}}, \bibinfo {author} {\bibfnamefont {Bohdan}\ \bibnamefont
  {{Kulchytskyy}}}, \bibinfo {author} {\bibfnamefont {Giacomo}\ \bibnamefont
  {{Torlai}}}, \ and\ \bibinfo {author} {\bibfnamefont {Roger~G.}\ \bibnamefont
  {{Melko}}},\ }\bibfield  {title} {\enquote {\bibinfo {title} {{The
  learnability scaling of quantum states: restricted Boltzmann machines}},}\
  }\href@noop {} {\bibfield  {journal} {\bibinfo  {journal} {arXiv e-prints}\
  ,\ \bibinfo {eid} {arXiv:1908.07532}} (\bibinfo {year} {2019})},\ \Eprint
  {http://arxiv.org/abs/1908.07532} {arXiv:1908.07532 [quant-ph]} \BibitemShut
  {NoStop}%
\bibitem [{\citenamefont {Luo}\ and\ \citenamefont {Clark}(2019)}]{backflowNN}%
  \BibitemOpen
  \bibfield  {author} {\bibinfo {author} {\bibfnamefont {Di}~\bibnamefont
  {Luo}}\ and\ \bibinfo {author} {\bibfnamefont {Bryan~K.}\ \bibnamefont
  {Clark}},\ }\bibfield  {title} {\enquote {\bibinfo {title} {Backflow
  transformations via neural networks for quantum many-body wave functions},}\
  }\href {\doibase 10.1103/PhysRevLett.122.226401} {\bibfield  {journal}
  {\bibinfo  {journal} {Phys. Rev. Lett.}\ }\textbf {\bibinfo {volume} {122}},\
  \bibinfo {pages} {226401} (\bibinfo {year} {2019})}\BibitemShut {NoStop}%
\bibitem [{\citenamefont {{Pfau}}\ \emph {et~al.}(2019)\citenamefont {{Pfau}},
  \citenamefont {{Spencer}}, \citenamefont {{Matthews}},\ and\ \citenamefont
  {{Foulkes}}}]{deepmindfermions}%
  \BibitemOpen
  \bibfield  {author} {\bibinfo {author} {\bibfnamefont {David}\ \bibnamefont
  {{Pfau}}}, \bibinfo {author} {\bibfnamefont {James~S.}\ \bibnamefont
  {{Spencer}}}, \bibinfo {author} {\bibfnamefont {Alexand er G. de~G.}\
  \bibnamefont {{Matthews}}}, \ and\ \bibinfo {author} {\bibfnamefont
  {W.~M.~C.}\ \bibnamefont {{Foulkes}}},\ }\bibfield  {title} {\enquote
  {\bibinfo {title} {{Ab-Initio Solution of the Many-Electron Schr{\"o}dinger
  Equation with Deep Neural Networks}},}\ }\href@noop {} {\bibfield  {journal}
  {\bibinfo  {journal} {arXiv e-prints}\ ,\ \bibinfo {eid} {arXiv:1909.02487}}
  (\bibinfo {year} {2019})},\ \Eprint {http://arxiv.org/abs/1909.02487}
  {arXiv:1909.02487 [physics.chem-ph]} \BibitemShut {NoStop}%
\bibitem [{\citenamefont {Torlai}\ and\ \citenamefont
  {Melko}(2018)}]{torlai_ndo}%
  \BibitemOpen
  \bibfield  {author} {\bibinfo {author} {\bibfnamefont {Giacomo}\ \bibnamefont
  {Torlai}}\ and\ \bibinfo {author} {\bibfnamefont {Roger~G.}\ \bibnamefont
  {Melko}},\ }\bibfield  {title} {\enquote {\bibinfo {title} {Latent space
  purification via neural density operators},}\ }\href {\doibase
  10.1103/PhysRevLett.120.240503} {\bibfield  {journal} {\bibinfo  {journal}
  {Phys. Rev. Lett.}\ }\textbf {\bibinfo {volume} {120}},\ \bibinfo {pages}
  {240503} (\bibinfo {year} {2018})}\BibitemShut {NoStop}%
\bibitem [{\citenamefont {Carrasquilla}\ \emph {et~al.}(2019)\citenamefont
  {Carrasquilla}, \citenamefont {Torlai}, \citenamefont {Melko},\ and\
  \citenamefont {Aolita}}]{carrasquilla_povm}%
  \BibitemOpen
  \bibfield  {author} {\bibinfo {author} {\bibfnamefont {Juan}\ \bibnamefont
  {Carrasquilla}}, \bibinfo {author} {\bibfnamefont {Giacomo}\ \bibnamefont
  {Torlai}}, \bibinfo {author} {\bibfnamefont {Roger~G.}\ \bibnamefont
  {Melko}}, \ and\ \bibinfo {author} {\bibfnamefont {Leandro}\ \bibnamefont
  {Aolita}},\ }\bibfield  {title} {\enquote {\bibinfo {title} {Reconstructing
  quantum states with generative models},}\ }\href {\doibase
  10.1038/s42256-019-0028-1} {\bibfield  {journal} {\bibinfo  {journal} {Nature
  Machine Intelligence}\ }\textbf {\bibinfo {volume} {1}},\ \bibinfo {pages}
  {155--161} (\bibinfo {year} {2019})}\BibitemShut {NoStop}%
\bibitem [{\citenamefont {{Glasser}}\ \emph {et~al.}(2019)\citenamefont
  {{Glasser}}, \citenamefont {{Sweke}}, \citenamefont {{Pancotti}},
  \citenamefont {{Eisert}},\ and\ \citenamefont
  {{Cirac}}}]{glasser2019expressive}%
  \BibitemOpen
  \bibfield  {author} {\bibinfo {author} {\bibfnamefont {Ivan}\ \bibnamefont
  {{Glasser}}}, \bibinfo {author} {\bibfnamefont {Ryan}\ \bibnamefont
  {{Sweke}}}, \bibinfo {author} {\bibfnamefont {Nicola}\ \bibnamefont
  {{Pancotti}}}, \bibinfo {author} {\bibfnamefont {Jens}\ \bibnamefont
  {{Eisert}}}, \ and\ \bibinfo {author} {\bibfnamefont {J.~Ignacio}\
  \bibnamefont {{Cirac}}},\ }\bibfield  {title} {\enquote {\bibinfo {title}
  {{Expressive power of tensor-network factorizations for probabilistic
  modeling, with applications from hidden Markov models to quantum machine
  learning}},}\ }\href@noop {} {\bibfield  {journal} {\bibinfo  {journal}
  {arXiv e-prints}\ ,\ \bibinfo {eid} {arXiv:1907.03741}} (\bibinfo {year}
  {2019})},\ \Eprint {http://arxiv.org/abs/1907.03741} {arXiv:1907.03741
  [cs.LG]} \BibitemShut {NoStop}%
\bibitem [{\citenamefont {Kokail}\ \emph {et~al.}(2019)\citenamefont {Kokail},
  \citenamefont {Maier}, \citenamefont {van Bijnen}, \citenamefont {Brydges},
  \citenamefont {Joshi}, \citenamefont {Jurcevic}, \citenamefont {Muschik},
  \citenamefont {Silvi}, \citenamefont {Blatt}, \citenamefont {Roos},\ and\
  \citenamefont {Zoller}}]{zoller19}%
  \BibitemOpen
  \bibfield  {author} {\bibinfo {author} {\bibfnamefont {C.}~\bibnamefont
  {Kokail}}, \bibinfo {author} {\bibfnamefont {C.}~\bibnamefont {Maier}},
  \bibinfo {author} {\bibfnamefont {R.}~\bibnamefont {van Bijnen}}, \bibinfo
  {author} {\bibfnamefont {T.}~\bibnamefont {Brydges}}, \bibinfo {author}
  {\bibfnamefont {M.~K.}\ \bibnamefont {Joshi}}, \bibinfo {author}
  {\bibfnamefont {P.}~\bibnamefont {Jurcevic}}, \bibinfo {author}
  {\bibfnamefont {C.~A.}\ \bibnamefont {Muschik}}, \bibinfo {author}
  {\bibfnamefont {P.}~\bibnamefont {Silvi}}, \bibinfo {author} {\bibfnamefont
  {R.}~\bibnamefont {Blatt}}, \bibinfo {author} {\bibfnamefont {C.~F.}\
  \bibnamefont {Roos}}, \ and\ \bibinfo {author} {\bibfnamefont
  {P.}~\bibnamefont {Zoller}},\ }\bibfield  {title} {\enquote {\bibinfo {title}
  {Self-verifying variational quantum simulation of lattice models},}\ }\href
  {\doibase 10.1038/s41586-019-1177-4} {\bibfield  {journal} {\bibinfo
  {journal} {Nature}\ }\textbf {\bibinfo {volume} {569}},\ \bibinfo {pages}
  {355--360} (\bibinfo {year} {2019})}\BibitemShut {NoStop}%
\bibitem [{\citenamefont {Carleo}\ \emph {et~al.}(2019)\citenamefont {Carleo},
  \citenamefont {Choo}, \citenamefont {Hofmann}, \citenamefont {Smith},
  \citenamefont {Westerhout}, \citenamefont {Alet}, \citenamefont {Davis},
  \citenamefont {Efthymiou}, \citenamefont {Glasser}, \citenamefont {Lin},
  \citenamefont {Mauri}, \citenamefont {Mazzola}, \citenamefont {Mendl},
  \citenamefont {van Nieuwenburg}, \citenamefont {O'Reilly}, \citenamefont
  {Th{\'e}veniaut}, \citenamefont {Torlai}, \citenamefont {Vicentini},\ and\
  \citenamefont {Wietek}}]{netket}%
  \BibitemOpen
  \bibfield  {author} {\bibinfo {author} {\bibfnamefont {Giuseppe}\
  \bibnamefont {Carleo}}, \bibinfo {author} {\bibfnamefont {Kenny}\
  \bibnamefont {Choo}}, \bibinfo {author} {\bibfnamefont {Damian}\ \bibnamefont
  {Hofmann}}, \bibinfo {author} {\bibfnamefont {James E.~T.}\ \bibnamefont
  {Smith}}, \bibinfo {author} {\bibfnamefont {Tom}\ \bibnamefont {Westerhout}},
  \bibinfo {author} {\bibfnamefont {Fabien}\ \bibnamefont {Alet}}, \bibinfo
  {author} {\bibfnamefont {Emily~J.}\ \bibnamefont {Davis}}, \bibinfo {author}
  {\bibfnamefont {Stavros}\ \bibnamefont {Efthymiou}}, \bibinfo {author}
  {\bibfnamefont {Ivan}\ \bibnamefont {Glasser}}, \bibinfo {author}
  {\bibfnamefont {Sheng-Hsuan}\ \bibnamefont {Lin}}, \bibinfo {author}
  {\bibfnamefont {Marta}\ \bibnamefont {Mauri}}, \bibinfo {author}
  {\bibfnamefont {Guglielmo}\ \bibnamefont {Mazzola}}, \bibinfo {author}
  {\bibfnamefont {Christian~B.}\ \bibnamefont {Mendl}}, \bibinfo {author}
  {\bibfnamefont {Evert}\ \bibnamefont {van Nieuwenburg}}, \bibinfo {author}
  {\bibfnamefont {Ossian}\ \bibnamefont {O'Reilly}}, \bibinfo {author}
  {\bibfnamefont {Hugo}\ \bibnamefont {Th{\'e}veniaut}}, \bibinfo {author}
  {\bibfnamefont {Giacomo}\ \bibnamefont {Torlai}}, \bibinfo {author}
  {\bibfnamefont {Filippo}\ \bibnamefont {Vicentini}}, \ and\ \bibinfo {author}
  {\bibfnamefont {Alexander}\ \bibnamefont {Wietek}},\ }\bibfield  {title}
  {\enquote {\bibinfo {title} {Netket: A machine learning toolkit for many-body
  quantum systems},}\ }\href {\doibase
  https://doi.org/10.1016/j.softx.2019.100311} {\bibfield  {journal} {\bibinfo
  {journal} {SoftwareX}\ }\textbf {\bibinfo {volume} {10}},\ \bibinfo {pages}
  {100311} (\bibinfo {year} {2019})}\BibitemShut {NoStop}%
\bibitem [{\citenamefont {Abraham}\ \emph {et~al.}(2019)\citenamefont
  {Abraham}, \citenamefont {Akhalwaya}, \citenamefont {Aleksandrowicz},
  \citenamefont {Alexander}, \citenamefont {Alexandrowics}, \citenamefont
  {Arbel}, \citenamefont {Asfaw}, \citenamefont {Azaustre}, \citenamefont
  {Barkoutsos}, \citenamefont {Barron}, \citenamefont {Bello}, \citenamefont
  {Ben-Haim}, \citenamefont {Bishop}, \citenamefont {Bosch}, \citenamefont
  {Bucher} \emph {et~al.}}]{qiskit}%
  \BibitemOpen
  \bibfield  {author} {\bibinfo {author} {\bibfnamefont {H{\'e}ctor}\
  \bibnamefont {Abraham}}, \bibinfo {author} {\bibfnamefont {Ismail~Yunus}\
  \bibnamefont {Akhalwaya}}, \bibinfo {author} {\bibfnamefont {Gadi}\
  \bibnamefont {Aleksandrowicz}}, \bibinfo {author} {\bibfnamefont {Thomas}\
  \bibnamefont {Alexander}}, \bibinfo {author} {\bibfnamefont {Gadi}\
  \bibnamefont {Alexandrowics}}, \bibinfo {author} {\bibfnamefont {Eli}\
  \bibnamefont {Arbel}}, \bibinfo {author} {\bibfnamefont {Abraham}\
  \bibnamefont {Asfaw}}, \bibinfo {author} {\bibfnamefont {Carlos}\
  \bibnamefont {Azaustre}}, \bibinfo {author} {\bibfnamefont {Panagiotis}\
  \bibnamefont {Barkoutsos}}, \bibinfo {author} {\bibfnamefont {George}\
  \bibnamefont {Barron}}, \bibinfo {author} {\bibfnamefont {Luciano}\
  \bibnamefont {Bello}}, \bibinfo {author} {\bibfnamefont {Yael}\ \bibnamefont
  {Ben-Haim}}, \bibinfo {author} {\bibfnamefont {Lev~S.}\ \bibnamefont
  {Bishop}}, \bibinfo {author} {\bibfnamefont {Samuel}\ \bibnamefont {Bosch}},
  \bibinfo {author} {\bibfnamefont {David}\ \bibnamefont {Bucher}},  \emph
  {et~al.},\ }\href {\doibase 10.5281/zenodo.2562110} {\enquote {\bibinfo
  {title} {Qiskit: An open-source framework for quantum computing},}\ }
  (\bibinfo {year} {2019})\BibitemShut {NoStop}%
\end{thebibliography}%


\section*{Supplementary Material}
In this Supplementary Material, we first discuss the standard technique to perform measurements of generic quantum observables in quantum hardware. Then we describe the representation of the many-body wavefunction with a restricted Boltzmann machine, and the approximate quantum state reconstruction technique. We also provide details on the methods used to generate the data shown in the manuscript.

\subsection*{Measurements in quantum hardware}
\label{SM}
A generic $N$-qubit quantum observable can be expressed as a linear combination 
\begin{equation}
\label{Observable}
\hat{\mathcal{O}}=\sum_{k=1}^K c_k \hat{P}_k,
\end{equation}
where $\hat{P}_k\in\{\hat{\mathbb{1}},\hat{\sigma}^x,\hat{\sigma}^y,\hat{\sigma}^z\}^{\otimes N}$ are tensor products of single-qubit Pauli operators. While direct measurement of eigen-states of the observable $\hat{\mathcal{O}}$ is in general not feasible, each Pauli operator $\hat{P}_k$ can be estimated independently on a quantum computer. Once a given quantum state $|\Psi\rangle$ of interest is prepared by the quantum hardware, $\hat{P}_k$ is measured by applying a suitable unitary transformation $\hat{\mathcal{U}}_k$ (compiled into a set of single-qubit gates) into the eigen-basis of $\hat{P}_k$, followed by single-qubit projective measurement. By measuring each Pauli operator in the expansion in Eq.~(\ref{Observable}) independently, an estimate for the observable $\hat{\mathcal{O}}$ is retrieved from a dataset $\mathcal{D}=\{\mathcal{D}_1,\dots,\mathcal{D}_K\}$. Each $\mathcal{D}_k=\{ {\bm\sigma}^{ {\bm k} }_1,\dots,\bm\sigma^{\bm k}_{S}\}$ is a collection of $S$ projective measurements $\bm\sigma^{\bm k}_j$ for the Pauli operator $\hat{P}_k$, where $\bm\sigma^{\bm k}_j$ is a $N$-bit string single-qubit measurement $\bm\sigma^{\bm k}_j = \{\sigma^{k_1}_{j,1},\dots, \sigma^{k_N}_{j,N}\}$ ($\sigma^{k_i}_{j,i} = \{0,1\}$) in the measurement basis ${\bm k}=(k_1,\dots,k_N)$ ($k_i=\{x,y,z\}$).

For simplicity, we assume in the following that the same number of measurements $S$ is used for each Pauli term, leading to $M=K\times S$ total queries to the quantum hardware. Given the measurement dataset $\mathcal{D}$, the expectation value of each Pauli operator and its variance are provided by the sample estimators
\begin{align}
\label{P_est}
\overline{ P}_k &= \sum_{j=1}^S \frac{P_{k,j}}{S}\\
\label{PV_est}
\sigma^2[P_k] &=\sum_{j=1}^S\frac{(P_{k,j}-\overline{P_k})^2}{S-1}\:,
\end{align}
where $P_{k,j} = \prod_{i=1}^N (-1)^{\sigma_{j,i}^{k_i}}$ is the result of a single measurement for the $k$-th Pauli operator. The standard way of building estimators for the observable $\hat{\mathcal{O}}$ on quantum computers follows from Eqs.~(\ref{P_est}) and (\ref{PV_est}):
\begin{align}
\overline{\mathcal{O}}_{\text{qc}} &= \sum_{k=1}^K c_k \overline{P}_k = \sum_{k=1}^K  \sum_{ j=1}^S \frac{c_kP_{k,j}}{S}\\
\sigma^2[\mathcal{O}]_{\text{qc}} &= \sum_{k=1}^K |c_k|^2 \sigma^2[P_k]  = \sum_{k=1}^K \sum_{j=1}^S \frac{|c_k|^2(P_{k,j}- \overline{ P}_k)^2}{S-1}\:.
\end{align}
The probability distribution of the measurement outcome $\overline{\mathcal{O}}_{\text{qc}}$ is a normal distribution with standard deviation 
\begin{equation}
\label{ErrMean}
\epsilon_{\text{qc}} = \frac{\sigma[\mathcal{O}]_{\text{qc}}}{\sqrt{S}}=\sqrt{ \sum_{k=1}^K \sum_{j=1}^S \frac{c_k^2(P_{k,j}- \overline{ P}_k)^2}{S(S-1)} }\:.
\end{equation}

A simple bound to this estimator is given by~\cite{wecker2015progress}
\begin{equation}
\epsilon^2_{\text{qc}} \le \epsilon^2_{\text{Max}} = \frac{(\sum_k|c_k|)^2}{M}\:.
\end{equation}
From this expression one can see how the error for this estimator is directly related to the number of Pauli operators $K$ in Eq.~(\ref{Observable}). Finally, we note that in the numerical simulation with synthetic data we have estimated the variance of the Pauli operators using the expectation value calculated on the exact ground state wavefunction, i.e. $\sigma^2[P_k] = 1-\langle\hat{P}_k\rangle^2$, rather than using sample variance from Eq.~(\ref{PV_est}).

\subsection*{Neural-network quantum reconstruction}
We propose to overcome the large measurement uncertainty of the standard procedure by using the measurement data to gain access to the quantum state underlying the hardware. Contrary to traditional quantum state tomography~\cite{Banaszek2013}, we perform an approximate reconstruction based on unsupervised learning of single-qubit projective measurement data with using artificial neural networks~\cite{torlai_2018_nnqst}. 

We adopt a representation of a pure quantum state based on a restricted Boltzmann machine (RBM), a stochastic neural network made out of two layers of binary units: a visible layer $\bm{\sigma}$ describing the qubits and a hidden layer $\bm{h}$, used to capture the correlations between the visible units~\cite{Hinton85}. The two layers are connected by a weight matrix $\bm{W}$, and additional fields (or biases) $\bm{c}$ and $\bm{d}$ couple to each unit in the two layers. Given a reference basis $|\bm{\sigma}\rangle=|\sigma_1,\dots,\sigma_N\rangle$ for $N$ qubits (e.g. $\sigma_j=\sigma_j^z$), the RBM provides the following (unnormalized) parametrization of the many-body wavefunction:
\begin{equation}
\begin{split}
\psi_{\bm{\lambda}}(\bm{\sigma})&=\sum_{\bm{h}}e^{\sum_{ij}W_{ij}\sigma_i h_j + \sum_j d_j h_j+\sum_i a_i\sigma_i}\\
&=e^{\sum_ia_i\sigma_i}e^{\sum_j\log\cosh\big(\sum_iW_{ij}\sigma_i+d_j\big)}\:.
\end{split}
\end{equation}
In order to capture quantum states with complex-valued amplitudes, we adopt complex-valued network parameters $\bm{\lambda}=\{\bm{a},\bm{W},\bm{d}\}$~\cite{Carleo17}. For a detailed description of the neural-network properties in the context of quantum many-body wavefunctions, we refer the reader to recent reviews~\cite{mlNISQ,RBM_natphys}.

The goal of the quantum state reconstruction is to discover a set of parameters $\bm{\lambda}$ such that the RBM wavefunction approximates an unknown target quantum state $\Psi$ on a set of measurement data. In a given measurement basis $\bm{b}=(b_1,\dots,b_N)$ ($b_i=\{x,y,z\}$) spanned by $|\bm{\sigma}^{\bm b}\rangle=|\sigma_1^{b_1},,\dots,\sigma_N^{b_N}\rangle$, the measurement probability distribution is  specified by the Born rule $P(\bm{\sigma}^{\bm b})=|\Psi(\bm{\sigma}^{\bm b})|^2$, where $\Psi(\bm{\sigma}^{\bm b})=\langle\bm{\sigma}^{\bm b}|\Psi\rangle$. Maximum likelihood learning of the network parameters corresponds to minimizing the extended Kullbach-Leibler (KL) divergence
\begin{equation}
\begin{split}
\mathcal{C}_{\bm{\lambda}}&=\sum_{\bm{b}}\sum_{\bm{\sigma}^{\bm b}}P(\bm{\sigma}^{\bm b})\log\frac{P(\bm{\sigma}^{\bm b})}{p_{\bm{\lambda}}(\bm{\sigma}^{\bf b})}\\
&\approx-\sum_{\bm{b}}\sum_{\bm{\sigma}^{\bm b}}P(\bm{\sigma}^{\bm b})\log p_{\bm{\lambda}}(\bm{\sigma}^{\bm b})\:,
\label{Eq::CostFull}
\end{split}
\end{equation} 
where the sum $\sum_{\bm{b}}$ runs on the informationally-complete set of $3^N$ bases and $\sum_{\bm{\sigma}^{\bm b}}$ runs over the full Hilbert space. We also omit the entropy contribution $\sum_{\bm{b}}\sum_{\bm{\sigma}^{\bm b}}P(\bm{\sigma}^{\bm b})\log P(\bm{\sigma}^{\bm b})$ since it does not depend on the parameters $\bm{\lambda}$. Note that $\mathcal{C}_{\bm{\lambda}}\ge0$, and $\mathcal{C}_{\bm{\lambda}}$ assumes its minimum value when $P(\bm{\sigma}^{\bm b})= p_{\bm{\lambda}}(\bm{\sigma}^{\bm b})$ $\forall\: \bm{b},\bm{\sigma}^{\bm b}$. Consequently, the optimal set of parameters $\bm{\lambda}^*=\argmin_{\bm{\lambda}}\mathcal{C}_{\bm{\lambda}}$ can be discovered by iterative updates of the form $\bm{\lambda}_{\omega}\rightarrow\bm{\lambda}_{\omega}-\eta\:\mathcal{G}_{\bm{\lambda}_{\omega}}$, where the {\it learning rate} $\eta$ is the size of the update, $\mathcal{G}_{\bm{\lambda}_{\omega}}=\frac{\partial}{\partial\bm{\lambda}_{\omega}}\mathcal{C}_{\bm{\lambda}}$ is the gradient of the cost function, and $\omega=\mathbb{R},\mathbb{I}$ indicates the real or imaginary part of the parameters and gradients.

In practice, the two exponentially large sums in Eq.~(\ref{Eq::CostFull}) are reduced using the finite-size training dataset $\mathcal{D}$, with measurement bases $\bm{b}$ corresponding to the set of Pauli operators $\hat{P}_k$ appearing in the decomposition of the observable in Eq.~(\ref{Observable}). This leads to the approximate negative-log likelihood (NLL):
\begin{equation}
\begin{split}
\mathcal{C}_{\bm{\lambda}}&=-\sum_{\bm{\sigma}^{\bm b}\in\mathcal{D}}\log p_{\bm{\lambda}}(\bm{\sigma}^{\bm b})\\
&=\log\sum_{\bm{\sigma}}|\psi_{\bm{\lambda}}(\bm{\sigma})|^2-
|\mathcal{D}|^{-1}\sum_{\bm{\sigma}^{\bm b}\in\mathcal{D}}\log|\psi_{\bm{\lambda}}(\bm{\sigma}^{\bm b})|^2\:,
\end{split}
\end{equation}
with $|\mathcal{D}|$ the size of the dataset. The RBM wavefunction in the $\bm{b}$ basis is
\begin{equation}
\psi_{\bm{\lambda}}(\bm{\sigma}^{\bm b})=\sum_{\bm{\sigma}}\mathcal{U}_{\bm{\sigma}^{\bm b}\bm{\sigma}}\psi_{\bm{\lambda}}(\bm{\sigma})\:,
\end{equation}
where $\hat{\mathcal{U}}$ is the unitary transformation that relates the measurement basis $|\bm{\sigma}^{\bm b}\rangle$ with the reference basis $|\bm{\sigma}\rangle$. As we restrict to the Pauli group, $\hat{\mathcal{U}}$ has a tensor product structure over each qubit, leading to the following matrix representation:
\begin{equation}
\mathcal{U}_{\bm{\sigma}^{\bm b}\bm{\sigma}}=
\prod_{j=1}^N\langle\sigma_j^{b_j}|\sigma_j\rangle\:.
\end{equation}

The gradient of the cost function $\mathcal{G}_{\bm{\lambda}}=\mathcal{G}_{{\bm{\lambda}_{\mathbb{R}}}}+i\mathcal{G}_{{\bm{\lambda}_{\mathbb{I}}}}$ can be calculated analytically:
\begin{equation}
\begin{split}
\mathcal{G}_{\bm{\lambda}_{\omega}}&=Z_{\bm{\lambda}}^{-1}\sum_{\bm{\sigma}}\frac{\partial}{\partial\bm{\lambda}^{\omega}}|\psi_{\bm{\lambda}}(\bm{\sigma})|^2\\
&\:-
|\mathcal{D}|^{-1}\sum_{\bm{\sigma}^{\bm b}\in\mathcal{D}}\frac{\partial}{\partial\bm{\lambda}^{\omega}}\log|\psi_{\bm{\lambda}}(\bm{\sigma}^{\bm b})|^2\\
&=Z_{\bm{\lambda}}^{-1}\sum_{\bm{\sigma}}|\psi_{\bm{\lambda}}(\bm{\sigma})|^2\frac{\partial}{\partial\bm{\lambda}^{\omega}}\Big(\log\psi_{\bm{\lambda}}(\bm{\sigma})+\log\psi^*_{\bm{\lambda}}(\bm{\sigma})\Big)+\\
&\:-|\mathcal{D}|^{-1}\sum_{\bm{\sigma}^{\bm b}\in\mathcal{D}}\frac{\partial}{\partial\bm{\lambda}^{\omega}}\Big(\log\psi_{\bm{\lambda}}(\bm{\sigma}^{\bm b})+\log\psi^*_{\bm{\lambda}}(\bm{\sigma}^{\bm b})\Big)\:.
\label{Eq::grad1}
\end{split}
\end{equation}
Here we have defined the wavefunction normalization $Z_{\bm{\lambda}}=\sum_{\bm{\sigma}}|\psi_{\bm{\lambda}}(\bm{\sigma})|^2$. The derivative of the wavefunction in the basis $\bm{b}$ is:
\begin{equation}
\begin{split}
\frac{\partial}{\partial\bm{\lambda}_{\omega}}\log\psi_{\bm{\lambda}}(\bm{\sigma}^{\bm b})&=
\frac{\sum_{\bm{\sigma}}\mathcal{U}_{\bm{\sigma}^{\bm b}\bm{\sigma}}\psi_{\bm{\lambda}}(\bm{\sigma})\frac{\partial}{\partial\bm{\lambda}_{\omega}}\log\psi_{\bm{\lambda}}(\bm{\sigma})}{\sum_{\bm{\sigma}}\mathcal{U}_{\bm{\sigma}^{\bm b}\bm{\sigma}}\psi_{\bm{\lambda}}(\bm{\sigma})}\\
&\equiv\big\langle\Phi_{\bm{\lambda}_{\omega}}(\bm{\sigma})\big\rangle_{\mathcal{Q}^{\bm{\sigma}_b}_{\bm{\lambda}}(\bm{\sigma})}
\label{Eq::DerLogRot}
\end{split}
\end{equation} 
where $\Phi_{\bm{\lambda}_{\omega}}(\bm{\sigma})=\frac{\partial}{\partial\bm{\lambda}_{\omega}}\log\psi_{\bm{\lambda}}(\bm{\sigma})$ and the average is taken with respect to the  quasi-probability distribution $\mathcal{Q}^{\bm{\sigma}_b}_{\bm{\lambda}}(\bm{\sigma})=\mathcal{U}_{\bm{\sigma}^{\bm b}\bm{\sigma}}\psi_{\bm{\lambda}}(\bm{\sigma})$. By inserting Eq.(~\ref{Eq::DerLogRot}) into the gradient in Eq.~(\ref{Eq::grad1}) we find
\begin{equation}
\mathcal{G}_{\bm{\lambda}_{\omega}}=2\big\langle\mathbb{R}\text{e}\big[\Phi_{\bm{\lambda}_{\omega}}(\bm{\sigma})\big]\big\rangle_{p_{\bm{\lambda}}(\bm{\sigma})}-2\big\langle\big\langle\mathbb{R}\text{e}\big[\Phi_{\bm{\lambda}_{\omega}}(\bm{\sigma})\big]\big\rangle_{\mathcal{Q}^{\bm b}_{\bm{\lambda}}(\bm{\sigma})}\big\rangle_{\mathcal{D}}
\end{equation}
The final gradient $\mathcal{G}_{\bm{\lambda}}$ can be written into a compact from by exploiting the holomorphic property of $\log\psi_{\bm{\lambda}}(\bm{\sigma})$. Following the definition of Wirtinger derivatives and applying the Cauchy-Riemann conditions,
\begin{equation}
\begin{split}
\Phi_{\bm{\lambda}}(\bm{\sigma})&=\frac{1}{2}(\Phi_{\bm{\lambda}_{\mathbb{R}}}(\bm{\sigma})-i\Phi_{\bm{\lambda},I}(\bm{\sigma}))\\
&=\mathbb{R}\text{e}\big[\Phi_{\bm{\lambda}_{\mathbb{R}}}(\bm{\sigma})\big]-i\mathbb{R}\text{e}\big[\Phi_{\bm{\lambda}_{\mathbb{I}}}(\bm{\sigma})\big],
\end{split}
\end{equation}
we can express the gradient as
\begin{equation}
\mathcal{G}_{\bm{\lambda}}=
2\Big[\big\langle\Phi_{\bm{\lambda}}^*(\bm{\sigma})\big\rangle_{p_{\bm{\lambda}}(\bm{\sigma})}-\big\langle\big\langle\Phi_{\bm{\lambda}}^*(\bm{\sigma})\big\rangle_{\mathcal{Q}^{\bm{\sigma}_b}_{\bm{\lambda}}(\bm{\sigma})}\big\rangle_{\mathcal{D}}\Big]\:.
\label{Eq::GradFinal}
\end{equation}

Similarly to a traditional RBM, the gradient breaks down into two components, traditionally called the \textit{positive} and \textit{negative} phase, driven respectively by the model and the data~\cite{Hinton85}. The negative-phase gradient is approximated by a Monte Carlo average
\begin{equation}
\big\langle\Phi_{\bm{\lambda}}^*(\bm{\sigma})\big\rangle_{p_{\bm{\lambda}}(\bm{\sigma})}
\approx\frac{1}{n}\sum_{i=1}^n\Phi_{\bm{\lambda}}^*(\bm{\sigma}_i)\:,
\end{equation}
where the configurations $\{\bm{\sigma}_i\}$ are sampled from the distribution $p_{\bm{\lambda}}(\bm{\sigma})$. The positive-phase gradient is averaged on the quasi-probability $\mathcal{Q}^{\bm{\sigma}_b}_{\bm{\lambda}}(\bm{\sigma})$
\begin{equation}
\big\langle\Phi_{\bm{\lambda}}^*(\bm{\sigma})\big\rangle_{\mathcal{Q}^{\bm{\sigma}_b}_{\bm{\lambda}}(\bm{\sigma})}=
\frac{\sum_{\bm{\sigma}}\mathcal{U}_{\bm{\sigma}^{\bm b}\bm{\sigma}}\psi_{\bm{\lambda}}(\bm{\sigma})\Phi_{\bm{\lambda}}^*(\bm{\sigma})}{\sum_{\bm{\sigma}}\mathcal{U}_{\bm{\sigma}^{\bm b}\bm{\sigma}}\psi_{\bm{\lambda}}(\bm{\sigma})}\:,
\end{equation}
which itself contains an intractable summation over the exponential size of the Hilbert space. However, the complexity of such expression can be reduced by careful choice of measurement bases. In fact, by measuring only a subset of $N_U$ qubits $\bm{\tau}=(\tau_1,\dots,\tau_{N_U})$ in local bases different than the reference one, the unitary rotation $\mathcal{U}$ simplifies to
\begin{equation}
\mathcal{U}_{\bm{\sigma}^{\bm b}\bm{\sigma}}=
\prod_{j=1}^{N_U}\langle\sigma_{\tau_j}^{b_{\tau_j}}|\sigma_{\tau_j}\rangle\prod_{\ell\notin\bm{\tau}}\delta_{\sigma_\ell^{b_{\ell}},\sigma_{\ell}}\:.
\end{equation}
where $\delta_{\alpha,\beta}$ is the Kronecker delta. By defining the basis $|\bm{s}\rangle=|s_1,\dots,s_{N_U}\rangle$ spanning the sub-space for the qubits being acted non-trivially upon by $\hat{\mathcal{U}}$ (i.e. $|s_j\rangle=|\sigma_{\tau_j}\rangle$), we obtain:
\begin{equation}
\sum_{\bm{\sigma}}\mathcal{U}_{\bm{\sigma}^{\bm b}\bm{\sigma}}\psi_{\bm{\lambda}}(\bm{\sigma})
=\sum_{\bm{s}}\prod_{j=1}^{N_U}\langle\sigma_{\tau_j}^{b_{\tau_j}}|\sigma_{\tau_j}\rangle
\Big[\bigotimes_{\ell\notin\bm{\tau}}\langle\sigma_\ell^z|\otimes\langle\bm{s}|\Big]|\psi_{\bm{\lambda}}\rangle\:,
\end{equation}
which now runs over $2^{N_U}$ terms. The scaling of the reconstruction algorithm is then $O(2^{N_U}n|\mathcal{D}|)$.

\paragraph{Neural-network estimator.} Once training is complete, the RBM is used to perform the measurement of the observable $\hat{\mathcal{O}}$. The expectation value of the observable on the RBM wavefunction is
\begin{equation}
\begin{split}
\frac{\langle\psi_{\bm{\lambda}}|\hat{\mathcal{O}}|\psi_{\bm{\lambda}}\rangle}{\langle\psi_{\bm{\lambda}}|\psi_{\bm{\lambda}}\rangle} &=Z_{\bm{\lambda}}^{-1}\sum_{\bm{\sigma}}|\psi_{\bm{\lambda}}(\bm{\sigma})|^2\frac{\langle\bm{\sigma}|\hat{\mathcal{O}}|\psi_{\bm{\lambda}}\rangle}{\langle\bm{\sigma}|\psi_{\bm{\lambda}}\rangle}\:.
\end{split}
\end{equation}
This can be approximated by Monte Carlo sampling, leading to the neural-network estimator
\begin{align}
\overline{\mathcal{O}}_{\bm{\lambda}}&=\frac{1}{n_{\text{mc}}}\sum_{j=1}^{n_{\text{mc}}}\mathcal{O}_{\bm{\lambda},j}\\
\sigma^2[\mathcal{O}]_{\bm{\lambda}} &= \sum_{j=1}^{n_{\text{mc}}} \frac{(\mathcal{O}_{\bm{\lambda},j}- \overline{\mathcal{O}}_{\bm{\lambda}})^2}{n_{\text{mc}}-1}\:,
\end{align}
where each single measurement is given by
\begin{equation}
\mathcal{O}_{\bm{\lambda},j}=\frac{\langle\bm{\sigma}_j|\hat{\mathcal{O}}|\psi_{\bm{\lambda}}\rangle}{\langle\bm{\sigma}_j|\psi_{\bm{\lambda}}\rangle}
=\sum_{\bm{\sigma}^\prime}\frac{\psi_{\bm{\lambda}}(\bm{\sigma}^\prime)}{\psi_{\bm{\lambda}}(\bm{\sigma})}\langle\bm{\sigma}|\hat{\mathcal{O}}|\bm{\sigma}^\prime\rangle\:,
\end{equation}
and $\bm{\sigma}_j$ are samples drawn from the neural-network distribution. Irrespective to the sampling procedure employed, the efficiency of the measurement procedure remains tied to the sparsity of the matrix representation of the observable $\hat{\mathcal{O}}$ in the reference basis. However, for most cases of interest, only a small number of matrix elements $\langle\bm{\sigma}|\hat{\mathcal{O}}|\bm{\sigma}^\prime\rangle$ are different from zero for a given $\bm{\sigma}$.

\paragraph{Training details.} In every training instance, we fix the total number of hidden units equal to the number of qubits $N$. Each sample $\bm{\sigma^b}$ in the training dataset is measured in a random basis uniformly sampled from the set of Pauli operators appearing in the observable decomposition. In practice, only a sub-set of the full training data (called {\it mini-batch}) is used to compute the gradient for a single update. The batch size was varied across training realizations, but  it never exceeded $10^4$. The parameters updates are carried out using the RMSprop optimizer, 
which introduces adaptive learning rates
\begin{equation}
\lambda^\prime_k=\lambda_k-\frac{\eta}{\sqrt{g_k}+\epsilon}\mathcal{G}_{\lambda_k}
\end{equation}
where the baseline value is set to $\eta=0.01$, $\epsilon=10^{-7}$ is a small off-set for numerical stability, and 	
\begin{equation}
g^\prime_k=\beta g_k+(1-\beta)\mathcal{G}^2_{\lambda_k}
\end{equation}
is the running average of the gradient squared ($\beta=0.9$). During both training and measurement process, the configurations $\{\bm{\sigma}_j\}$ sampled from the RBM distribution, required respectively for calculating the negative phase and the neural-network estimator, are obtained using parallel tempering with 20-25 parallel chains.

In order to select the optimal set of RBM parameters $\bm{\lambda}$, we split the dataset $\mathcal{D}$ and use 90$\%$ of data for training and the remaining 10$\%$ for validation. During training, we evaluate the NLL on the validation set, and save a fixed number of different network parameters generating the lowest values (usually between 100 and 500). For the case of synthetic data, the best set of parameters $\bm{\lambda}^*$ is selected (among this set) as the one the generates the lowest energy. For experimental data, we keep the set that generates the lowest NLL on the validation set. Finally, we note that the calculation of the NLL requires the knowledge of the partition function $Z_{\bm{\lambda}}$, which was computed exactly in the numerical experiments. In general, an approximation to $Z_{\bm{\lambda}}$ (and thus to NLL) can be obtained using either parallel tempering or annealed importance sampling.

\end{document}